\documentclass[11pt]{article}
\usepackage{amssymb}
\usepackage{colortbl}
\usepackage{amsfonts,amsmath, longtable}

\usepackage{comment}

\usepackage{color}

\newtheorem{corollary}{Corollary}[section]

        \def\theequation{\thesection.\arabic{equation}}

\newcommand{\tr}{{\rm tr}}
\newcommand{\ti}[1]{\tilde{#1}}

\newcommand{\mH}{{\mathcal H}}

\newcommand{\vth}{\vartheta}

\newcommand{\Mat}{ {\rm Mat}(N,\mathbb C) }

\newcommand{\mC}{\mathbb C}
\newcommand{\mZ}{\mathbb Z}

\newcommand{\z}{{\zeta}}

\newcommand{\tiL}{{\tilde L}}

\newtheorem{predl}{Proposition}[section]

\def\beq{\begin{equation}}
\def\eq{\end{equation}}
\def\p{\partial}

\newtheorem{theor}{Theorem}[section]

\def\res{\mathop{\hbox{Res}}\limits}

\begin{document}

\setcounter{page}{1}

\begin{center}

\

\vspace{0mm}

{\Large{\bf  Classical $r$-matrix structure for elliptic Ruijsenaars chain and }}

\vspace{3mm}

{\Large{\bf
1+1 field analogue of Ruijsenaars-Schneider model
}}



 \vspace{15mm}

 {\Large {D. Murinov}}
\qquad\quad\quad
 {\Large {A. Zotov}}\footnote{Corresponding Author.}

  \vspace{10mm}



 {\em Steklov Mathematical Institute of Russian
Academy of Sciences,\\ Gubkina str. 8, 119991, Moscow, Russia}

   \vspace{3mm}

{\em Institute for Theoretical and Mathematical Physics,\\
Lomonosov Moscow State University, Moscow, 119991, Russia}

 {\small\rm {e-mails: murinov344@yandex.ru, zotov@mi-ras.ru}}

\end{center}

\vspace{0mm}

\begin{abstract}
The classical dynamical $r$-matrix structure for the periodic elliptic Ruijsenaars chain is described.
The Poisson brackets for the monodromy matrix are calculated as well, thus providing
Liouville integrability of the model. Next, we study its continuous non-relativistic limit
and reproduce the Maillet type non-ultralocal $r$-matrix structure for the field analogue
of the elliptic Calogero-Moser model.
\end{abstract}

%

\bigskip

{\small{ \tableofcontents }}

\bigskip

\section{Introduction and overview}\label{sec1}
\setcounter{equation}{0}
\paragraph{Many-body systems.}In this paper, we deal with the Ruijsenaars-Schneider \cite{RS} and the Calogero-Moser \cite{Calogero2} models --
two widely known elliptic many-body integrable systems.
The first one is described by the Hamiltonian
  \beq\label{q01}
  \begin{array}{l}
  \displaystyle{
 H^{\hbox{\tiny{RS}}}=
 c\sum\limits_{j=1}^N \, \prod\limits_{k:k\neq j}^N\frac{\vth(q_{j}-q_k-\eta)}{\vth(q_{j}-q_k)}\,
 e^{p_j/c}\,,
 }
 \end{array}
 \eq
where $p_i$ and $q_i$, $i=1,...,N$ are canonically conjugated momenta and positions of particles
on a complex plane, $c\in\mC$ and $\eta\in\mC$ are constants and $\vth(w)$ is the first Jacobi theta-function
(\ref{a03}). In the non-relativistic limit $c\to\infty$ (together with redefining $\eta=\nu/c$) one gets
the elliptic Calogero-Moser model:
  \beq\label{q02}
  \begin{array}{c}
  \displaystyle{
H^{\hbox{\tiny{CM}}}=\sum\limits_{i=1}^N\frac{p_i^2}{2}-\nu^2\sum\limits_{i>j}^N\wp(q_i-q_j)\,,
 }
 \end{array}
 \eq
where $\wp(w)$ is the Weierstrass $\wp$-function (\ref{a072}).

Both models are described through the Lax representation
  \beq\label{q03}
  \begin{array}{c}
  \displaystyle{
 \dot{L}(z)+[L(z),M(z)]=0\,,\qquad L(z),M(z)\in\Mat
 }
 \end{array}
 \eq
with the spectral parameter $z$. Integrability follows from the existence of the classical
$r$-matrix structure, which guarantees the involution property $\{\tr L^k(z), \tr L^m(w)\}=0$.
The $r$-matrix structure for the Calogero-Moser model is linear \cite{BradenSuz,Skl3}
  \beq\label{q04}
  \begin{array}{c}
  \displaystyle{
\{L^{\hbox{\tiny{CM}}}_1(z),L^{\hbox{\tiny{CM}}}_2(w)\}=
[L^{\hbox{\tiny{CM}}}_1(z),r^{\hbox{\tiny{CM}}}_{12}(z,w)]-[L^{\hbox{\tiny{CM}}}_2(w),r^{\hbox{\tiny{CM}}}_{21}(w,z)]
 }
 \end{array}
 \eq
 and it is quadratic for the Ruijsenaars-Schneider model \cite{Suris,NKSR}:
\beq\label{q05}
  \begin{array}{c}
  \displaystyle{
c\left\{L^{\hbox{\tiny{RS}}}_1(z),L^{\hbox{\tiny{RS}}}_2(w)\right\} =L^{\hbox{\tiny{RS}}}_1(z)L^{\hbox{\tiny{RS}}}_2(w)r_{12}^{-}(z,w) -
r_{12}^{+}(z,w)L^{\hbox{\tiny{RS}}}_1(z)L^{\hbox{\tiny{RS}}}_2(w) +
 }
 \\ \ \\
   \displaystyle{
+
L^{\hbox{\tiny{RS}}}_1(z)s_{12}^+(z,w)L^{\hbox{\tiny{RS}}}_2(w)-
L^{\hbox{\tiny{RS}}}_2(w)s_{12}^-(z,w)L^{\hbox{\tiny{RS}}}_1(z)\,.
 }
 \end{array}
 \eq
Explicit expressions for all $r$-matrices are given in the main part of the paper.

\paragraph{Chains.}The periodic classical
elliptic Ruijsenaars chain was proposed in \cite{ZZ}. It is a generalization of the
Ruijsenaars-Schneider model, which has $n$ sites. That is, the phase space $\mC^{2Nn}$ is parameterized
by $Nn$ pairs of canonical variables $p^a_i$, $q_i^a$ $i=1,...,N$, $a=1,...,n$:
  \beq\label{q06}
  \begin{array}{c}
  \displaystyle{
 \{p^a_i,q_j^b\}=\delta^{ab}\delta_{ij}\,,\quad
 \{p^a_i,p_j^b\}=\{q^a_i,q_j^b\}=0\,,\quad i,j=1,...,N;\ a,b=1,...,n\,.
 }
 \end{array}
 \eq
The numeration is cyclic assuming  $q^0_i= q^n_i$, $p^0_i= p^n_i$ and
 $q^{n+1}_i= q^1_i$, $p^{n+1}_i= p^1_i$.
The Hamiltonian has the form
\beq\label{q07}
 \begin{array}{c}
  \displaystyle{
 H=c\sum\limits_{a=1}^n \log h_{a,a+1}\,,\qquad
  h_{a-1,a}=\sum\limits_{j=1}^N
 \frac{\prod\limits_{l=1}^N\vth({ q}^a_j-{ q}^{a-1}_l-\eta) }
 {\vth(-\eta)\prod\limits_{l: l\neq j}^N\vth({ q}^{a}_j-{ q}^{a}_l) }\,e^{p^a_j/c}\,.
 }
 \end{array}
  \eq
 The chain is described by a set of $n$ Lax matrices $L^a\in\Mat$ satisfying the lattice semi-discrete
 zero-curvature (Zakharov-Shabat) equation
\beq\label{q08}
 \begin{array}{c}
  \displaystyle{
  {{\dot L}^a}(z)=\{H,{ L}^a(z)\}={ L}^a(z){ M}^a(z)-{ M}^{a-1}(z){ L}^a(z)\,,\quad a=1,...,n
 }
 \end{array}
  \eq
Then the monodromy matrix
  \beq\label{q09}
  \begin{array}{c}
  \displaystyle{
 T(z)=L^1(z)L^2(z)...L^n(z)\in\Mat
 }
 \end{array}
 \eq
satisfies the Lax equation
  \beq\label{q10}
  \begin{array}{c}
  \displaystyle{
 {\dot T}(z)=[T(z),M^n(z)]\,.
 }
 \end{array}
 \eq
In fact, the first elliptic model of type (\ref{q07}) was proposed by I. Krichever in \cite{KrToda} and
called the elliptic Toda chain. 
Another model of this type is the Ruijsenaars-Toda chain
introduced by V. Adler and Yu. Suris in \cite{Adler}.
The precise connection between (\ref{q07}) and these models requires several additional steps, which we will not need in this article. We will describe this relation in a separate publication.

The Ruijsenaars chain was derived in \cite{ZZ} in two ways. The first one is by gauge transformation of the
XYZ chain, and the second way is by certain reduction from 2d Toda hierarchy. The original Ruijsenaars-Schneider model
can also be obtained from 2d Toda lattice as well \cite{KrZ}, and its hyperbolic version is obtained from the sine-Grodon
model (two-periodic reduction from 2d Toda) \cite{Babelon,Pogr}. Below, we consider the non-relativistic limit, assuming the chain is relativistic.
This terminology is taken from the original papers \cite{RS}, where the term "relativistic" arose
due to the existence of the Poincar\'e algebra. In fact, we do not know if the chain model admits the Poincar\'e algebra.
We use the term "relativistic" to refer to (group-like) quadratic (in $L$) $r$-matrix structures, and the term "non-relativistic" is used to refer to models described by linear (Lie algebra-like) $r$-matrix structures.

\paragraph{Field theories.} It was shown in \cite{Krich2,LOZ,Krich22} that the Calogero-Moser model (\ref{q02})
can be extended to an integrable 1+1 field theory. In this case the dynamical variables become periodic functions
$p_i(x)$, $q_i(x)$ on a unit circle $p_i(x)=p_i(x+2\pi)$, $q_i(x)=q_i(x+2\pi)$ with the canonical Poisson brackets
 \beq\label{q11}
 \begin{array}{c}
  \displaystyle{
\{p_i(x),q_j(y)\}=\delta_{ij}\delta(x-y)\,,\qquad \{p_i(x),p_j(y)\}=\{q_i(x),q_j(y)\}=0\,.
  }
 \end{array}
\eq
The Hamiltonian (\ref{q02}) is generalized to
 \beq\label{q12}
 \begin{array}{c}
  \displaystyle{
  \mH^{\hbox{\tiny{2dCM}}}=\int\limits_{0}^{2\pi} d x H^{\hbox{\tiny{2dCM}}}(x)
  }
 \end{array}
\eq
with the density
 \beq\label{q13}
 \begin{array}{c}
  \displaystyle{
H^{\hbox{\tiny{2dCM}}}(x)=-\sum\limits_{j=1}^N p_j^2(kq_{j,x}+\nu)+\frac{1}{N\nu}\Big(\sum\limits_{j=1}^N p_j(kq_{j,x}+\nu)\Big)^2
+\frac14\sum\limits_{j=1}^N\frac{k^4 q_{j,xx}^2}{kq_{j,x}+\nu}+
  }
  \\
    \displaystyle{
    +\frac12\sum\limits_{i\neq j}^N\Big( (kq_{i,x}+\nu)^2(kq_{j,x}+\nu)+(kq_{i,x}+\nu)(kq_{j,x}+\nu)^2
    -\nu k^2(q_{i,x}-q_{j,x})^2  \Big)\wp(q_i-q_j)+
      }
  \\
    \displaystyle{
    +\frac{k^3}{2}\sum\limits_{i\neq j}^N\Big( q_{i,x}q_{j,xx}-q_{i,xx}q_{j,x} \Big)\zeta(q_i-q_j)\,,
    }
 \end{array}
\eq
where $q_{i,x}=\p_xq_i(x)$ and $q_{i,xx}=\p^2_xq_i(x)$, and $\zeta(w)$ is the Weierstrass zeta-function (\ref{a071}).
The limit $k\to 0$ corresponds to vanishing of all $\p_x$--derivatives, that is to the limit to finite-dimensional
classical mechanics. In this limit $H^{\hbox{\tiny{2dCM}}}(x)\to -2\nu H^{\hbox{\tiny{CM}}}$.
This model is described by the Zakharov-Shabat equation
\cite{ZSh}:
 \beq\label{q14}
 \begin{array}{c}
  \displaystyle{
 \partial_{t}{U}^{\hbox{\tiny{2dCM}}}(z)-k\partial_{x}{V}^{\hbox{\tiny{2dCM}}}(z)
 +[{U}^{\hbox{\tiny{2dCM}}}(z), {V}^{\hbox{\tiny{2dCM}}}(z)]=0
  }
 \end{array}
\eq
 with certain U-V (or L-A) pair
${U}^{\hbox{\tiny{2dCM}}}(z)\,,{V}^{\hbox{\tiny{2dCM}}}(z)\in\Mat$. The infinite-dimensional analogue of the
Liouville integrability is as follows. Consider the monodromy matrix
 \beq\label{q15}
 \begin{array}{c}
  \displaystyle{
 T(z,x)={\rm Pexp}\Big( \frac{1}{k}\int\limits_0^x dy\,  U(z,y) \Big)\,.
  }
 \end{array}
\eq
It was shown in \cite{Maillet} that the Poisson commutativity $\{\tr(T^k(z,2\pi)),\tr(T^m(w,2\pi))\}=0$
follows from the existence of the Maillet $r$-matrix structure
  \beq\label{q16}
  \begin{array}{c}
  \displaystyle{
\{U_1(z,x),U_2(w,y)\}=
  }
  \\ \ \\
  \displaystyle{
\Big(-k\p_x {\bf r}_{12}(z,w|x)
+ [U_1(z,x),{\bf r}_{12}(z,w|x)]-[U_2(w,y),{\bf r}_{21}(w,z|x)]\Big)\delta(x-y)-
  }
  \\ \ \\
  \displaystyle{
  -\Big({\bf r}_{12}(z,w|x)+{\bf r}_{21}(w,z|x)\Big)k\delta'(x-y)\,,
 }
 \end{array}
 \eq
 and this relation for 1+1 Calogero-Moser field theory $U(z,x)=U^{\hbox{\tiny{2dCM}}}(z,x)$
 was proved in \cite{Z24}.

A wide class of Hitchin-type integrable systems was generalized
 to the field level during the last years \cite{DZ,Z24spin,ZZ}, including
the Ruijsenaars-Schneider model. The field analogue of the Hamiltonian  (\ref{q01}) is a straightforward
 field analogue of its lattice version (\ref{q07}), see \cite{ZZ}:
\beq\label{q18}
 \begin{array}{c}
  \displaystyle{
 \mH^{\hbox{\tiny{2dRS}}}=\int\limits_{0}^{2\pi} d x H^{\hbox{\tiny{2dRS}}}(x)\,,\qquad
 H^{\hbox{\tiny{2dRS}}}(x)=c \log \sum\limits_{j=1}^N
 \frac{\prod\limits_{l=1}^N\vth({ q}_j(x)-{ q}_l(x-\eta)-\eta) }
 {\vth(-\eta)\prod\limits_{l: l\neq j}^N\vth({ q}_j(x)-{ q}_l(x)) }\,e^{p_j(x)/c}\,.
 }
 \end{array}
  \eq
Instead of the lattice equation (\ref{q08}) in the field case, we deal with the semi-discrete
 Zakharov-Shabat equation
\beq\label{q19}
 \begin{array}{c}
  \displaystyle{
  {\p_t U}^{\hbox{\tiny{2dRS}}}(z,x)=
  U^{\hbox{\tiny{2dRS}}}(z,x)V^{\hbox{\tiny{2dRS}}}(z,x)-
  V^{\hbox{\tiny{2dRS}}}(z,x-\eta)U^{\hbox{\tiny{2dRS}}}(z,x)\,.
 }
 \end{array}
  \eq

\paragraph{Purpose of the paper.} The first goal of this paper is to describe $r$-matrix structure
of the Ruijsenaars chain (\ref{q06})-(\ref{q09}). Specifically, we compute the Poisson brackets
$\{L^a_1(z),L^b_2(w)\}$ and represent them in $r$-matrix form similarly to the original Ruijsenaars-Schneider
model (\ref{q05}) corresponding to the case $n=1$. This allows us to compute the brackets for the monodromy
matrix $\{T_1(z),T_2(w)\}$, and these results imply the Poisson commutativity
 $\{\tr T^k(z),\tr T^l(w)\}=0$.
The second goal is to show that the obtained $r$-matrix structure generalizes the Maillet non-ultralocal
brackets (\ref{q16}) for the field analogue of the Calogero-Moser model (\ref{q11})-(\ref{q13}). To do this, we move to the field theory level (\ref{q18})-(\ref{q19}) and calculate the continuous non-relativistic limit.
We show that the brackets (\ref{q16}) are indeed reproduced in this limit.

The paper is organized as follows. We first recall the description of the original Ruijsenaars-Schneider model and its
$r$-matrix structure together with the non-relativistic limit in Section \ref{sec2}. Next, we describe the
classical $r$-matrix structure for the Ruijsenaars chain in Section \ref{sec3}. In Section \ref{sec4}
we proceed to the field theory description and then perform the limit to the Maillet brackets in Section \ref{sec5}.
Necessary elliptic functions definitions and properties are given in the Appendix A. In the Appendix B
the original description from \cite{NKSR} of the $r$-matrix structure of the Ruijsenaars-Schneider model
is recalled since the one we use slightly differs from it.


\section{Elliptic Ruijsenaars-Schneider model}\label{sec2}
\setcounter{equation}{0}

\subsection{Hamiltonian description and Lax pair}
The classical $N$-body elliptic Ruijsenaars-Schneider model is described by the
following Lax matrix
  \beq\label{q101}
  \begin{array}{l}
  \displaystyle{
 L^{\hbox{\tiny{RS}}}_{ij}(z)=\phi(z,q_{i}-q_j+\eta)\,b_j\,,\ i,j=1,\ldots ,N\,,
 }
 \end{array}
 \eq
  \beq\label{q102}
  \begin{array}{l}
  \displaystyle{
 b_j=\prod_{k:k\neq j}^N\frac{\vth(q_{j}-q_k-\eta)}{\vth(q_{j}-q_k)}\,
 e^{p_j/c}\,,
 }
 \end{array}
 \eq
defined through the Kronecker elliptic function $\phi(z,u)$ (\ref{a01}) and the first Jacobi theta function $\vth(z)$ (\ref{a03}). The parameters $\eta, c\in\mC$ are constants. Positions of particles $q_j$ and momenta $p_i$
are canonically conjugated variables
 %
  \beq\label{q103}
  \begin{array}{c}
    \displaystyle{
\{p_i,q_j\}=\delta_{ij}\,,\quad \{p_i,p_j\}=\{q_i,q_j\}=0\,.
 }
 \end{array}
 \eq
The Hamiltonian comes from evaluating the trace of the Lax matrix (\ref{q101}):
  \beq\label{q104}
  \begin{array}{l}
  \displaystyle{
 H^{\hbox{\tiny{RS}}}=c\frac{\tr L^{\hbox{\tiny{RS}}}(z)}{\phi(z,\eta)}=c\sum\limits_{j=1}^N b_j=
 c\sum\limits_{j=1}^N \, \prod\limits_{k:k\neq j}^N\frac{\vth(q_{j}-q_k-\eta)}{\vth(q_{j}-q_k)}\,
 e^{p_j/c}\,.
 }
 \end{array}
 \eq
 The Hamiltonian equations of motion take the form
  \beq\label{q105}
  \begin{array}{l}
  \displaystyle{
 {\dot q}_j=\{H^{\hbox{\tiny{RS}}},q_j\}=\p_{p_j}H^{\hbox{\tiny{RS}}}=b_j\,,
 }
 \end{array}
 \eq
 that is
  \beq\label{q106}
  \begin{array}{l}
  \displaystyle{
 L^{\hbox{\tiny{RS}}}_{ij}(z)=\phi(z,q_{i}-q_j+\eta)\,{\dot q}_j\,,\quad H^{\hbox{\tiny{RS}}} =c\sum\limits_{j=1}^N{\dot q}_j(p,q)\,.
 }
 \end{array}
 \eq
 The second set of Hamiltonian equations is
  \beq\label{q107}
  \begin{array}{c}
  \displaystyle{
 {\dot p}_i=\{H^{\hbox{\tiny{RS}}},p_i\}=-\,\p_{q_i}H^{\hbox{\tiny{RS}}}
  =c\sum\limits_{l:l\neq i}^N ({\dot q}_i+{\dot q}_l)E_1(q_{il})-{\dot q}_i E_1(q_{il}-\eta)
  -{\dot q}_l E_1(q_{il}+\eta)\,,
  }
 \end{array}
 \eq
 where $q_{ij}=q_i-q_j$ and $E_1(z)$ is the first Eisenstein function (\ref{a07}).
 In the  Newtonian form we have:
 \beq\label{q108}
  \begin{array}{c}
  \displaystyle{
 {\ddot q}_i=\sum\limits_{k:k\neq i}^N{\dot q}_i{\dot q}_k
 (2E_1(q_{ik})-E_1(q_{ik}+\eta)-E_1(q_{ik}-\eta))\,,\quad i=1, \ldots ,
 N\,.
 }
 \end{array}
 \eq
The equations of motion (\ref{q108}) are represented in the Lax form
  \beq\label{q109}
  \begin{array}{c}
  \displaystyle{
 {\dot L}^{\hbox{\tiny{RS}}}(z)\equiv\{H^{\hbox{\tiny{RS}}},L^{\hbox{\tiny{RS}}}(z)\}=[L^{\hbox{\tiny{RS}}}(z),M^{\rm
 RS}(z)]
 }
 \end{array}
 \eq
  with the accompany $M$-matrix
  \beq\label{q110}
  \begin{array}{c}
  \displaystyle{
 M^{\hbox{\tiny{RS}}}_{ij}(z)=
 -(1-\delta_{ij})\phi(z,q_i-q_j)\,{\dot q}_j-
 }
 \\ \ \\
   \displaystyle{
 -\delta_{ij}\Big({\dot q}_i\,(E_1(z)+E_1(\eta)) +
 \sum\limits_{k:k\neq i}^N {\dot q}_k\,(E_1(q_{ik}+\eta)-E_1(q_{ik}))
 \Big)\,,\ i,j=1,\ldots ,N\,.
 }
 \end{array}
 \eq
This statement is verified by direct calculation using identities (\ref{a08})-(\ref{a11}).

\subsection{Classical $r$-matrix}
The classical $r$-matrix structure for the elliptic Ruijsenaars-Schneider model was
proposed in \cite{NKSR}. The form of the Lax matrix in  \cite{NKSR} differs from
 (\ref{q101}) by substitution of $b_j\rightarrow b_i$, that is these two Lax matrices
are related by the gauge transformation with the diagonal gauge transformation matrix ${\rm diag}(b_1,...,b_N)$.
For this reason, the answer given below for the $r$-matrix structure differs slightly from the original.
For the reader's convenience, we recall the original r-matrix from \cite{NKSR}
along with some standard notation in Appendix B.

\begin{predl}\label{prop1}
For the matrix $L^{\hbox{\tiny{RS}}}(z)$ (\ref{q101}) the following quadratic $r$-matrix relation holds:
\beq\label{q121}
  \begin{array}{c}
  \displaystyle{
c\left\{L^{\hbox{\tiny{RS}}}_1(z),L^{\hbox{\tiny{RS}}}_2(w)\right\} =L^{\hbox{\tiny{RS}}}_1(z)L^{\hbox{\tiny{RS}}}_2(w)r_{12}^{-}(z,w) -
r_{12}^{+}(z,w)L^{\hbox{\tiny{RS}}}_1(z)L^{\hbox{\tiny{RS}}}_2(w) +
 }
 \\ \ \\
   \displaystyle{
+
L^{\hbox{\tiny{RS}}}_1(z)s_{12}^+(z,w)L^{\hbox{\tiny{RS}}}_2(w)-
L^{\hbox{\tiny{RS}}}_2(w)s_{12}^-(z,w)L^{\hbox{\tiny{RS}}}_1(z)\,,
 }
 \end{array}
 \eq
where
\begin{equation}\label{q122}
   \displaystyle{
    r_{12}^- (z,w) = r_{12}(z,w)-s^+_{12}(z,w)+s^-_{12}(z,w)\,,
    }
\end{equation}
\begin{equation}\label{q123}
    r_{12}^+(z,w) = r_{12}(z,w)\,,
\end{equation}
\begin{equation}\label{q124}
    s_{12}^+ (z,w) = s_{12}(z)+u_{12}^+\,,
\end{equation}
\begin{equation}\label{q125}
    s_{12}^- (z,w) = s_{21}(w)-u_{12}^-
\end{equation}
with
\begin{equation}\label{q126}
    r_{12}(z,w) = \sum_{i\neq j}^N\phi(z-w,q_{ij})E_{ij}\otimes E_{ji}+E_1(z-w)\sum_{i=1}^N E_{ii}\otimes E_{ii} - \sum_{i\neq j}^{N} E_1(q_{ij})E_ {ii}\otimes E_{jj}\,,
\end{equation}
\begin{equation}\label{q127}
    u_{12}^{\pm} = \sum\limits_{i,j=1}^N E_1(q_{ji}\pm \eta)E_{ii}\otimes E_{jj}\,,
\end{equation}
and
\begin{equation}\label{q128}
    s_{12}(z) = \sum_{i,j,m=1}^N (L^{\hbox{\tiny{RS}}}(z)^{-1})_{mi}(''\partial_{\eta}L^{\hbox{\tiny{RS}}}(z)'')_{ij} E_{mj}\otimes E_{ii}\,.
\end{equation}
In the last definition $''\partial_{\eta}L^{\hbox{\tiny{RS}}}(z)''$ means that the derivative with respect to $\eta$
acts on the function $\phi$ as in (\ref{a08}) but does not act on $b_j$:
\begin{equation}\label{q129}
''\partial_{\eta}L^{\hbox{\tiny{RS}}}(z)''_{ij}=L^{\hbox{\tiny{RS}}}_{ij}(z)\Big(E_1(z+q_{ij}+\eta)-E_1(q_{ij}+\eta)\Big)\,.
\end{equation}
That is
\begin{equation}\label{q130}
    L^{\hbox{\tiny{RS}}}_1(z)s_{12}(z) = \sum_{i,j=1}^N L^{\hbox{\tiny{RS}}}_{ij}(z)
    \Big(E_1(z+q_{ij}+\eta)-E_1(q_{ij}+\eta)\Big)E_{ij}\otimes E_{ii}\,,
\end{equation}
and similarly
\begin{equation}\label{q131}
    L^{\hbox{\tiny{RS}}}_2(w)s_{21}(w) = \sum_{i,j=1}^N
    L^{\hbox{\tiny{RS}}}_{ij}(w)\Big(E_1(w+q_{ij}+\eta)-E_1(q_{ij}+\eta)\Big)E_{ii}\otimes E_{ij}\,.
\end{equation}
\end{predl}
This statement and its proof are very similar to one from \cite{NKSR}. It is verified directly
using (\ref{a08})-(\ref{a11}) for all tensor components $E_{ij}\otimes E_{kl}$
including all possibilities of coinciding indices.

The conditions (\ref{b05}) and (\ref{b07}) are fulfilled due to
\begin{equation}\label{q132}
    r_{21}^{\pm}(w,z) = -r_{12}^{\pm}(z,w), \quad s_{21}^+(w,z) = s_{12}^-(z,w);
\end{equation}
\begin{equation}\label{q133}
    r_{12}^+(z,w)-s_{12}^+(z,w) = r_{12}^-(z,w)-s_{12}^-(z,w).
\end{equation}
Therefore,
\beq\label{q134}
  \begin{array}{c}
  \displaystyle{
\{\tr \Big(L^{\hbox{\tiny{RS}}}(z)\Big)^k,\tr \Big(L^{\hbox{\tiny{RS}}}(w)\Big)^m\} =0\,.
}
 \end{array}
 \eq

Finally, let us mention that the definition of the Lax matrix (\ref{q101}) has some freedom generated
by the canonical map
  \beq\label{q135}
  \begin{array}{c}
  \displaystyle{
p_j\ \rightarrow\ p_j+c_1\log \prod\limits_{k:k\neq j}^N
\frac{\vth(q_{j}-q_k+\eta)}{\vth(q_{j}-q_k-\eta)}
 }
 \end{array}
 \eq
with an arbitrary constant $c_{1}$. For example, by choosing $c_1=c$ the functions $b_j$ are changed as
  \beq\label{q136}
  \begin{array}{c}
  \displaystyle{
b_j\ \rightarrow\  \prod\limits_{k:k\neq j}^N
\frac{\vth(q_{j}-q_k+\eta)}{\vth(q_{j}-q_k)}\,e^{p_j/c}\,,
 }
 \end{array}
 \eq
and for $c_1=c/2$ we have
  \beq\label{q137}
  \begin{array}{c}
  \displaystyle{
b_j\ \rightarrow\  \prod\limits_{k:k\neq j}^N
\frac{\Big(\vth(q_{j}-q_k+\eta)\vth(q_{j}-q_k+\eta)\Big)^{1/2}}{\vth(q_{j}-q_k)}\,e^{p_j/c}
\stackrel{(\ref{a12})}{=}
  }
  \\
  \displaystyle{
=
\left(\frac{\vth(\eta)}{\vth'(0)}\right)^{N-1}\prod\limits_{k:k\neq j}^N
\Big(\wp(\eta)-\wp(q_{j}-q_k)\Big)^{1/2}\,e^{p_j/c}\,,
 }
 \end{array}
 \eq
and this form with square root was used originally in \cite{RS}.
Since the canonical maps preserve the Poisson brackets $\{b_i,b_j\}$ and $\{b_i,q_j\}$
the $r$-matrix structure changes (\ref{q121}) by replacing functions $b_j$ with their
transformed versions. That is, all
the Lax matrices $L^{\hbox{\tiny{RS}}}$ entering the r.h.s. of (\ref{q121}) should be taken
with the transformed $b_j$ as well.

\subsection{Non-relativistic limit to Calogero-Moser model}

\paragraph{The limit in the Lax matrix.} Introduce notation
  \beq\label{q201}
  \begin{array}{c}
  \displaystyle{
\varepsilon=\frac{1}{c}\,,\qquad \eta=\frac{\nu}{c}=\varepsilon\nu\,.
 }
 \end{array}
 \eq
Then the Lax matrix (\ref{q101})-(\ref{q102}) takes the form
  \beq\label{q202}
  \begin{array}{c}
  \displaystyle{
 L^{\hbox{\tiny{RS}}}_{ij}(z)=\phi(z,q_{i}-q_j+\varepsilon\nu)\,b_j\,,\quad
 b_j=\prod_{k:k\neq j}^N\frac{\vth(q_{j}-q_k-\varepsilon\nu)}{\vth(q_{j}-q_k)}\,
 e^{\varepsilon p_j}\,.
 }
 \end{array}
 \eq
The non-relativistic limit corresponds to $c\rightarrow\infty$, that is to $\varepsilon\rightarrow 0$.
Due to (\ref{a06}) we have the following expansion near $\varepsilon=0$:
  \beq\label{q203}
  \begin{array}{c}
  \displaystyle{
\phi(z,q_{i}-q_j+\varepsilon\nu)=\frac{1}{\varepsilon\nu}\,\delta_{ij}
+\Big( \delta_{ij}E_1(z)+(1-\delta_{ij})\phi(z,q_{i}-q_j) \Big)+O(\varepsilon)\,,
 }
 \end{array}
 \eq
and for $b_j$
  \beq\label{q204}
  \begin{array}{c}
  \displaystyle{
b_j=1+\varepsilon{\ti p}_j+O(\varepsilon^2)\,,\quad {\ti p}_j=p_j-\nu\sum\limits_{k\neq j}^NE_1(q_j-q_k)\,.
 }
 \end{array}
 \eq
Notice that $p_j\to {\ti p}_j$ is a canonical map. In this way, we obtain
  \beq\label{q205}
  \begin{array}{c}
  \displaystyle{
 L^{\hbox{\tiny{RS}}}(z)=\frac{1}{\varepsilon\nu}\,1_N+\frac{1}{\nu}\,L^{\hbox{\tiny{CM}}}(z)+O(\varepsilon)\,,
 }
 \end{array}
 \eq
where $L^{\hbox{\tiny{CM}}}(z)$ is the Lax matrix for the elliptic Calogero-Moser model \cite{Krich1}
with the spectral parameter $z$ and the coupling constant $\nu$:
  \beq\label{q206}
  \begin{array}{c}
  \displaystyle{
 L^{\hbox{\tiny{CM}}}_{ij}(z)=\delta_{ij}\Big({\ti p}_i+\nu E_1(z)\Big)+(1-\delta_{ij})\nu\phi(z,q_{i}-q_j)\,.
 }
 \end{array}
 \eq

\paragraph{The limit in $r$-matrix structure.}
Consider the limit for the $r$-matrix structure (\ref{q121}).
For this purpose, it is useful to write (\ref{q121}) in the following form:
\beq\label{q210}
  \begin{array}{c}
  \displaystyle{
c\left\{L^{\hbox{\tiny{RS}}}_1(z),L^{\hbox{\tiny{RS}}}_2(w)\right\} =[L^{\hbox{\tiny{RS}}}_1(z)L^{\hbox{\tiny{RS}}}_2(w),r_{12}(z,w)] +
 }
 \\ \ \\
   \displaystyle{
+
[L^{\hbox{\tiny{RS}}}_1(z),L^{\hbox{\tiny{RS}}}_2(w)s_{12}^-(z,w)]-
[L^{\hbox{\tiny{RS}}}_2(w),L^{\hbox{\tiny{RS}}}_1(z)s_{12}^+(z,w)]
\,.
 }
 \end{array}
 \eq
In the limit $\eta\to 0$, its l.h.s.  behaves as
  \beq\label{q211}
  \begin{array}{c}
  \displaystyle{
c\{L^{\hbox{\tiny{RS}}}_1(z),L^{\hbox{\tiny{RS}}}_2(w)\}=\frac{1}{\varepsilon\nu^2}\,\{L^{\hbox{\tiny{CM}}}_1(z),L^{\hbox{\tiny{CM}}}_2(w)\}+O(1)\,.
 }
 \end{array}
 \eq
The expression $r_{12}(z,w)$ (\ref{q321}) is independent of $\eta$. Therefore,
  \beq\label{q212}
  \begin{array}{c}
  \displaystyle{
[L^{\hbox{\tiny{RS}}}_1(z)L^{\hbox{\tiny{RS}}}_2(w),r_{12}(z,w)]=\frac{1}{\nu\eta}\,[L^{\hbox{\tiny{CM}}}_1(z)+L^{\hbox{\tiny{CM}}}_2(w),r_{12}(z,w)]+O(1)=
 }
 \\ \ \\
   \displaystyle{
=\frac{1}{\varepsilon\nu^2}\,[L^{\hbox{\tiny{CM}}}_1(z),r_{12}(z,w)]-\frac{1}{\varepsilon\nu^2}\,[L^{\hbox{\tiny{CM}}}_2(w),r_{21}(w,z)]+O(1)\,,
 }
 \end{array}
 \eq
where the last equality came from the skew-symmetry property $r_{12}(z,w)=-r_{21}(w,z)$.

Next, consider the limit for the second line of (\ref{q210}). It is verified directly that
all the terms with singularities of orders $1/\varepsilon^3$ and $1/\varepsilon^2$ are cancelled out.
Let us compute the order $1/\varepsilon$, which provides an input into the final answer:
\beq\label{q2121}
  \begin{array}{c}
  \displaystyle{
\res\limits_{\eta=0}\Big([L^{\hbox{\tiny{RS}}}_1(z),L^{\hbox{\tiny{RS}}}_2(w)s_{12}^-(z,w)]-
[L^{\hbox{\tiny{RS}}}_2(w),L^{\hbox{\tiny{RS}}}_1(z)s_{12}^+(z,w)]\Big)=
 }
 \\ \ \\
   \displaystyle{
=\frac{1}{\nu}\,[L^{\hbox{\tiny{CM}}}_1(z),\res\limits_{\eta=0}\Big(L^{\hbox{\tiny{RS}}}_2(w)s_{12}^-(z,w)\Big)]-
\frac{1}{\nu}\,[L^{\hbox{\tiny{CM}}}_2(w),\res\limits_{\eta=0}\Big(L^{\hbox{\tiny{RS}}}_1(z)s_{12}^+(z,w)\Big)]\,.
 }
 \end{array}
 \eq
For $L^{\hbox{\tiny{RS}}}_2(w)s_{12}^-(z,w)$ we have
  \beq\label{q213}
  \begin{array}{c}
  \displaystyle{
-L^{\hbox{\tiny{RS}}}_2(w)u^-_{12}=\sum\limits_{i,j,k=1}^N L^{\hbox{\tiny{RS}}}_{kj}(w)E_1(q_{ij}+\eta) E_{ii}\otimes E_{kj}=
\sum\limits_{k=1}^N\sum\limits_{i\neq j}^N L^{\hbox{\tiny{RS}}}_{kj}(w)E_1(q_{ij}+\eta) E_{ii}\otimes E_{kj}+
 }
  \\
   \displaystyle{
   +\sum\limits_{k\neq j}^N L^{\hbox{\tiny{RS}}}_{kj}(w)E_1(\eta)E_{jj}\otimes E_{kj}+\sum\limits_{j=1}^N L^{\hbox{\tiny{RS}}}_{jj}(w)E_1(\eta)E_{jj}\otimes E_{jj}
 }
 \end{array}
 \eq
and
  \beq\label{q214}
  \begin{array}{c}
  \displaystyle{
L^{\hbox{\tiny{RS}}}_2(w)s_{21}(w)\stackrel{(\ref{q131})}{=}
\sum\limits_{i\neq j}^N L^{\hbox{\tiny{RS}}}_{ij}(w)\Big(E_1(w+q_{ij}+\eta)-E_1(q_{ij}+\eta)\Big) E_{ii}\otimes E_{ij}+
 }
  \\
   \displaystyle{
   +\sum\limits_{j=1}^N L^{\hbox{\tiny{RS}}}_{jj}(w)\Big(E_1(w+\eta)-E_1(\eta)\Big) E_{jj}\otimes E_{jj}\,.
 }
 \end{array}
 \eq
To summarize the above, we see that the term with $E_1(\eta)L^{\hbox{\tiny{RS}}}_{jj}(w)$ cancels:
  \beq\label{q215}
  \begin{array}{c}
  \displaystyle{
L^{\hbox{\tiny{RS}}}_2(w)s^-_{12}(z,w)=
\sum\limits_{k=1}^N\sum\limits_{i\neq j}^N L^{\hbox{\tiny{RS}}}_{kj}(w)E_1(q_{ij}+\eta) E_{ii}\otimes E_{kj}+
 }
  \\
   \displaystyle{
   +\sum\limits_{k\neq j}^N L^{\hbox{\tiny{RS}}}_{kj}(w)E_1(\eta)E_{jj}\otimes E_{kj}+\sum\limits_{j=1}^N L^{\hbox{\tiny{RS}}}_{jj}(w)E_1(w+\eta)E_{jj}\otimes E_{jj}\,.
 }
 \end{array}
 \eq
Therefore,
  \beq\label{q216}
  \begin{array}{c}
  \displaystyle{
\res\limits_{\eta=0}\Big(L^{\hbox{\tiny{RS}}}_2(w)s_{12}^-(z,w)\Big)=
 }
  \\
   \displaystyle{
=\sum\limits_{i\neq j}^N E_1(q_{ij}) E_{ii}\otimes E_{jj}+
\frac{1}{\nu}\sum\limits_{i\neq j}^N L^{\hbox{\tiny{CM}}}_{ij}(w)E_{jj}\otimes E_{ij}
+E_1(w)\sum\limits_{i=1}^N E_{ii}\otimes E_{ii}\,.
 }
 \end{array}
 \eq
Notice that
  \beq\label{q2161}
  \begin{array}{c}
  \displaystyle{
\frac{1}{\nu}\sum\limits_{i\neq j}^N L^{\hbox{\tiny{CM}}}_{ij}(w)E_{jj}\otimes E_{ij}
=\sum\limits_{i\neq j}^N \phi(w,q_{ij})E_{jj}\otimes E_{ij}=
-\sum\limits_{i\neq j}^N \phi(-w,q_{ij})E_{ii}\otimes E_{ji}\,.
 }
 \end{array}
 \eq
The calculation for the expression $L^{\hbox{\tiny{RS}}}_1(z)s_{12}^+(z,w)$ is similar:
  \beq\label{q217}
  \begin{array}{c}
  \displaystyle{
\res\limits_{\eta=0}\Big(L^{\hbox{\tiny{RS}}}_1(z)s_{12}^+(z,w)\Big)=
 }
  \\
   \displaystyle{
=\sum\limits_{i\neq j}^N E_1(q_{ij}) E_{jj}\otimes E_{ii}+
\frac{1}{\nu}\sum\limits_{i\neq j}^N L^{\hbox{\tiny{CM}}}_{ij}(z)E_{ij}\otimes E_{jj}
+E_1(z)\sum\limits_{i=1}^N E_{ii}\otimes E_{ii}\,.
 }
 \end{array}
 \eq
Plugging it into (\ref{q2121}) together with (\ref{q211})-(\ref{q212}) this yields
  \beq\label{q218}
  \begin{array}{c}
  \displaystyle{
\{L^{\hbox{\tiny{CM}}}_1(z),L^{\hbox{\tiny{CM}}}_2(w)\}=
[L^{\hbox{\tiny{CM}}}_1(z),r^{\hbox{\tiny{CM}}}_{12}(z,w)]-[L^{\hbox{\tiny{CM}}}_2(w),r^{\hbox{\tiny{CM}}}_{21}(w,z)]\,,
 }
 \end{array}
 \eq
where
\beq\label{q219}
  \begin{array}{c}
  \displaystyle{
 r^{\hbox{\tiny{CM}}}_{12}(z,w)=r_{12}(z,w)+\res\limits_{\eta=0}\Big(L^{\hbox{\tiny{RS}}}_2(w)s_{12}^-(z,w)\Big)
 =(E_1(z-w)+E_1(w))\,\sum\limits_{i=1}^NE_{ii}\otimes E_{ii}+
  }
  \\
   \displaystyle{
 +
 \sum\limits^N_{i\neq j}\phi(z-w,q_{ij})\,E_{ij}\otimes E_{ji}
 -\sum\limits^N_{i\neq j}\phi(-w,q_{ij})\,E_{ii}\otimes E_{ji}
 }
 \end{array}
 \eq
 is the classical $r$-matrix for the elliptic Calogero-Moser model \cite{Skl3,BradenSuz}.

\section{Finite-dimensional case: Ruijsenaars chain}\label{sec3}
\setcounter{equation}{0}

\subsection{Description of the model}
The periodic ${\rm GL}_N$ elliptic Ruijsenaars chain on $n$ sites is defined as follows \cite{ZZ}.
Its phase space $\mC^{2Nn}$ is parameterized by canonical coordinates
  \beq\label{q301}
  \begin{array}{c}
  \displaystyle{
 \{p^a_i,q_j^b\}=\delta^{ab}\delta_{ij}\,,\quad
 \{p^a_i,p_j^b\}=\{q^a_i,q_j^b\}=0\,,\quad i,j=1,...,N;\ a,b=1,...,n\,.
 }
 \end{array}
 \eq
The numeration of sites (the upper indices) is modulo $n$. By definition,
we assume
  \beq\label{q302}
  \begin{array}{c}
  \displaystyle{
 q^0_i= q^n_i\,,\ p^0_i= p^n_i\,,\quad
 q^{n+1}_i= q^1_i\,,\ p^{n+1}_i= p^1_i\,,\quad i=1,...,N\,.
 }
 \end{array}
 \eq
Introduce the monodromy matrix
  \beq\label{q303}
  \begin{array}{c}
  \displaystyle{
 T(z)=L^1(z)L^2(z)...L^n(z)\in\Mat
 }
 \end{array}
 \eq
with the Lax matrices
 \beq\label{q304}
 \begin{array}{c}
  \displaystyle{
 L^a_{ij}(z)=
 \phi(z,{ q}^{a-1}_i-{ q}^{a}_j+\eta)
 \frac{\prod\limits_{l=1}^N\vth({ q}^a_j-{ q}^{a-1}_l-\eta) }
 {\vth(-\eta)\prod\limits_{l: l\neq j}^N\vth({ q}^{a}_j-{ q}^{a}_l) }\,e^{p^a_j/c}\,,
 \quad a=1,...,n;\quad i,j=1,...,N\,.
 }
 \end{array}
 \eq
The Hamiltonian is calculated as
\beq\label{q305}
 \begin{array}{c}
  \displaystyle{
 \exp(H/c)=\res\limits_{z=0}z^{n-1}\tr T(z)\,.
 }
 \end{array}
  \eq
  Then
\beq\label{q306}
 \begin{array}{c}
  \displaystyle{
 H=c\sum\limits_{a=1}^n \log h_{a,a+1}
 }
 \end{array}
  \eq
 with
   \beq\label{q307}
 \begin{array}{c}
  \displaystyle{
 h_{a-1,a}=\sum\limits_{j=1}^N b^a_j=\sum\limits_{j=1}^N
 \frac{\prod\limits_{l=1}^N\vth({ q}^a_j-{ q}^{a-1}_l-\eta) }
 {\vth(-\eta)\prod\limits_{l: l\neq j}^N\vth({ q}^{a}_j-{ q}^{a}_l) }\,e^{p^a_j/c}\,.
 }
 \end{array}
  \eq
  It provides the following equations of motion (in the Newtonian form):
   \beq\label{q308}
 \begin{array}{c}
  \displaystyle{
  \frac{{\ddot q}^a_i }{ {\dot q}^a_i }=
   -\sum\limits_{l=1}^N {\dot { q}}_l^{a+1}E_1({ q}_i^a-{ q}_l^{a+1}+\eta)
  -\sum\limits_{l=1}^N {\dot { q}}_l^{a-1}E_1({ q}_i^a-{ q}_l^{a-1}-\eta)
  +2\sum\limits_{l\neq i}^N {\dot {q}}_l^{a}E_1({q}_i^a-{q}_l^{a})+
 }
 \\
   \displaystyle{
   +\sum\limits_{m,l=1}^N {\dot { q}}_m^{a}{\dot { q}}_l^{a+1}E_1({ q}_m^a-{ q}_l^{a+1}+\eta)
  -\sum\limits_{m,l=1}^N {\dot { q}}_l^{a}{\dot { q}}_m^{a-1}E_1({ q}_m^{a-1}-{ q}_l^a+\eta)\,.
   }
 \end{array}
  \eq
These equations are represented in the form of Lax equations, or more precisely, in
the form of semi-discrete Zakharov-Shabat equation:
\beq\label{q309}
 \begin{array}{c}
  \displaystyle{
  {{\dot L}^a}(z)=\{H,{ L}^a(z)\}={ L}^a(z){ M}^a(z)-{ M}^{a-1}(z){ L}^a(z)\,.
 }
 \end{array}
  \eq
Details of the zero-curvature representation can be found in \cite{ZZ}.

\subsection{Classical $r$-matrix structure}
Here we formulate main results.
\begin{theor}\label{th1}
 The Lax matrices $L^a(z)$ (\ref{q304}) satisfy the following quadratic $r$-matrix structure:
\begin{equation}\label{q320}
 \begin{array}{c}
   \displaystyle{
     c\{L_{1}^a(z),L_2^b(w)\}=
     \delta^{ab}\Big(L_1^b(z)L_2^b(w)r^b_{12}(z,w)-r_{12}^{b-1}(z,w)L_1^b(z)L_2^b(w)+
   }
     \\ \ \\
   \displaystyle{
     +L_1^b(z)s^{+,\,b}_{12}(z)L_2^b(w)-L_2^b(w)s_{12}^{-,\,b}(w)L_1^b(z)\Big)+
   }
     \\ \ \\
   \displaystyle{
     +\delta^{a,\,b-1}L_1^{b-1}(z)L_2^b(w)s_{12}^{-,\,b}(w)-
     \delta^{a,\,b+1}L_1^{b+1}(z)L_2^b(w)s_{12}^{+,\, b+1}(z)\,,
   }
 \end{array}
\end{equation}
where
\begin{equation}\label{q321}
 \begin{array}{c}
   \displaystyle{
    r_{12}^a(z,w)=
       }
     \\ \ \\
   \displaystyle{
    =\sum\limits_{i\neq j}^N\phi(z-w,q_i^a-q_j^a)E_{ij}\otimes E_{ji}
    +E_1(z-w)\sum\limits_{i=1}^N E_{ii}\otimes E_{ii}
    -\sum\limits_{i\neq j}^N E_1(q_i^a-q_j^a) E_{ii}\otimes E_{jj}\,,
    }
  \end{array}
\end{equation}
\begin{equation}\label{q322}
    s_{12}^{+,\,a}(z)=s_{12}^a(z)+u_{12}^{+,\,a}\,,
\end{equation}
\begin{equation}\label{q323}
    s_{12}^{-,\,a}(w)=s_{21}^a(w)-u_{12}^{-,\,a}\,,
\end{equation}
\begin{equation}\label{q324}
    u_{12}^{+,\,a}=\sum\limits_{i,j=1}^NE_1(q_j^{a-1}-q_i^a+\eta)E_{ii}\otimes E_{jj}\,,
\end{equation}
\begin{equation}\label{q325}
    u_{12}^{-,\,a}=-\sum\limits_{i,j=1}^NE_1(q_i^{a-1}-q_j^a+\eta)E_{ii}\otimes E_{jj}\,,
\end{equation}
and the matrices $s_{12}^a(z)$ are defined similarly to (\ref{q130}) through
\begin{equation}\label{q326}
    L_1^a(z)s_{12}^a(z)=\sum\limits_{i,j=1}^N L_{ij}^a(z)
    \Big(E_1(z+q_i^{a-1}-q_j^a+\eta)-E_1(q_i^{a-1}-q_{j}^a+\eta)\Big)E_{ij}\otimes E_{ii}\,,
\end{equation}
\begin{equation}\label{q327}
    L_2^a(w)s_{21}^a(w)=\sum\limits_{i,j=1}^N L_{ij}^a(w)
    \Big(E_1(w+q_i^{a-1}-q_j^a+\eta)-E_1(q_i^{a-1}-q_{j}^a+\eta)\Big)E_{ii}\otimes E_{ij}\,.
\end{equation}
The delta-symbols $\delta^{a,\,b-1}$ and $\delta^{a,\,b+1}$ in (\ref{q320}) are understood modulo $n$
according to
(\ref{q302}).
\end{theor}
The proof is similar to the one from \cite{NKSR} and Proposition \ref{prop1}.

Next, we use the relations (\ref{q320}) to derive $r$-matrix structure for the
monodromy matrix $T(z)$ (\ref{q303}).

\begin{theor}\label{th2}
The monodromy matrix $T(z)$ (\ref{q303}) satisfies the following $r$-matrix structure:
\begin{equation}\label{q330}
 \begin{array}{c}
   \displaystyle{
     \{T_1(z),T_2(w)\}=T_1(z)T_2(w)r_{12}^n(z,w)-\left(r_{12}^n(z,w)-
     \breve{s}_{12}^{+,1}(z,w)+\breve{s}_{12}^{-,1}(z,w)\right)T_1(z)T_2(w)
       }
     \\ \ \\
   \displaystyle{
   +T_1(z)\breve{s}_{12}^{-,1}(z,w)T_2(w)-T_2(w)\breve{s}_{12}^{+,1}(z,w)T_1(z)\,,
    }
  \end{array}
\end{equation}
where $r_{12}^n(z,w)$ is defined in (\ref{q321}) for $a=n$, and $\breve{s}_{12}^{\pm,1}(z,w)$
are obtained by conjugation
\begin{equation}\label{q331}
 \begin{array}{c}
   \displaystyle{
    \breve{s}_{12}^{+,1}(z,w)=L_1^1(z)s_{12}^{+,1}(z,w)\left(L_1^1(z)\right)^{-1}\,,
       }
     \\ \ \\
   \displaystyle{
    \breve{s}_{12}^{-,1}(z,w)=L_2^1(w)s_{12}^{-,1}(z,w)\left(L_2^1(w)\right)^{-1}
    }
  \end{array}
\end{equation}
from ${s}_{12}^{\pm,1}(z,w)$ defined in (\ref{q322})-(\ref{q323}) for $n=1$.
\end{theor}

The proof is by direct substitution. By writing down $\{T_1(z),T_2(w)\}$ through the Leibnitz rule
and plugging (\ref{q320}), one gets (\ref{q330}).

Let us comment on the form of matrices $\breve{s}_{12}^{\pm,1}(z,w)$. The
conjugation (\ref{q331}) of ${s}_{12}^{\pm,1}(z,w)$ (\ref{q322})-(\ref{q323}) yields
\begin{equation}\label{q332}
 \begin{array}{c}
   \displaystyle{
    \breve{s}_{12}^{+,1}(z,w)=\breve{s}_{12}^1(z)+\breve{u}_{12}^{+,1}(z)\,,
           }
     \\ \ \\
   \displaystyle{
   \breve{s}_{12}^{-,1}(z,w)=\breve{s}_{21}^1(w)-\breve{u}_{12}^{-,1}(w)\,,
    }
  \end{array}
\end{equation}
where with the notation (\ref{q129}) we have
\begin{equation}\label{q333}
    \breve{s}_{12}^1(z)=L_1^1(z){s}_{12}^1(z)\left(L_1^1(z)\right)^{-1}
    =
    \sum\limits_{i,j=1}^N\left(''\partial_\eta L^1(z)''\left(L^{1}(z)\right)^{-1}\right)_{ij}E_{ij}\otimes E_{ii}
\end{equation}
and similarly
\begin{equation}\label{q334}
    \breve{s}_{21}^1(w)=L_2^1(w){s}_{21}^1(w)\left(L_2^1(w)\right)^{-1}
    =
    \sum\limits_{i,j=1}^N\left(''\partial_\eta L^1(w)''\left(L^{1}(w)\right)^{-1}\right)_{ij}E_{ii}\otimes E_{ij}\,.
\end{equation}
For the matrices $\breve{u}_{12}^{\pm,1}$ one gets
\begin{equation}\label{q335}
    \breve{u}_{12}^{+,1}(z)=L_1^1(z){u}_{12}^{+,1}(z)\left(L_1^1(z)\right)^{-1}=
    \sum_{i,j,k,l=1}^NL^1_{ki}(z)\left(L^1(z)\right)^{-1}_{il}E_1\left(q_j^n-q_i^1+\eta\right)E_{kl}\otimes E_{jj}\,,
\end{equation}
\begin{equation}\label{q336}
    \breve{u}_{12}^{-,1}(w)=
    L_2^1(w){u}_{12}^{-,1}\left(L_2^1(w)\right)^{-1}
    =-\sum_{i,j,k,l=1}^NL^1_{kj}(w)\left(L^1(w)\right)^{-1}_{jl}E_1\left(q_i^n-q_j^1+\eta\right)E_{ii}\otimes E_{kl}\,.
\end{equation}

It is easy to verify that the properties (\ref{b05}) for the brackets (\ref{q330}) (being written in the form (\ref{b01}))
hold true. Therefore, we come to the following statement.
\begin{corollary}
It follows from (\ref{q330}) that
\beq\label{q337}
  \begin{array}{c}
  \displaystyle{
\{\tr \left(T^k(z)\right),\tr \left(T^l(w)\right)\} =0\,.
}
 \end{array}
 \eq
\end{corollary}
%



\section{Field theory}\label{sec4}
\setcounter{equation}{0}

\subsection{1+1 Ruijsenaars-Schneider field theory}
In this section, we use the following notations:
\begin{equation}\label{q521}
    \dfrac{1}{c}=-\varepsilon; \quad \eta_{rel} =-\nu\varepsilon; \quad \eta_{chain}=\varepsilon,
\end{equation}
where $\eta_{rel}$ -- parameter of relativistic deformation and $\eta_{chain}$ -- period of the chain. The non-relativistic limit corresponds to $c\to\infty$ or $\varepsilon \to 0$. It is performed similarly to the limit
for the Ruijsenaars-Schneider model described in
Section ~\ref{sec2}.
In the introduced notation, the $U$-matrix for the Ruijsenaars-Schneider field theory takes the form:
\begin{equation}\label{q522}
  \begin{array}{c}
  \displaystyle{
        U_{ij}(z,x)=\phi(z,q_i(x-\varepsilon)-q_j(x)-\nu\varepsilon)
        \frac{\prod\limits_{a=1}^N\vartheta(q_j(x)-q_a(x-\varepsilon)+\nu\varepsilon)}{\vartheta(\nu\varepsilon)
        \prod\limits_{a:a\neq j}^N\vartheta(q_j(x)-q_a(x))}e^{-\varepsilon p_j(x)}\,.
  }
  \end{array}
\end{equation}
The equations of motion are generated by the redefined Hamiltonian~\eqref{q18}:
\beq\label{q5231}
 \begin{array}{c}
  \displaystyle{
 \mH^{\hbox{\tiny{2dRS}}}=\int\limits_{0}^{2\pi} d x H^{\hbox{\tiny{2dRS}}}(x)\,,\qquad
 H^{\hbox{\tiny{2dRS}}}(x)=-\dfrac{1}{\varepsilon} \log \sum\limits_{j=1}^N
 \frac{\prod\limits_{l=1}^N\vth({ q}_j(x)-{ q}_l(x-\varepsilon)+\nu\varepsilon) }
 {\vth(\nu\varepsilon)\prod\limits_{l: l\neq j}^N\vth({ q}_j(x)-{ q}_l(x)) }\,e^{-\varepsilon p_j(x)}\,.
 }
 \end{array}
  \eq
Let us rewrite it in the form
\begin{equation}\label{q5241}
    \begin{gathered}
        H^{\hbox{\tiny{2dRS}}}(x)=-\dfrac{1}{\varepsilon} \log h(x)\,,
    \end{gathered}
\end{equation}
where
\begin{equation}\label{q5251}
    h(x)=\sum\limits_{j=1}^N
 \frac{\prod\limits_{l=1}^N\vth({ q}_j(x)-{ q}_l(x-\varepsilon)+\nu\varepsilon) }
 {\vth(\nu\varepsilon)\prod\limits_{l: l\neq j}^N\vth({ q}_j(x)-{ q}_l(x)) }\,e^{-\varepsilon p_j(x)}.
\end{equation}
Then the Hamiltonian equations are of the form~\cite{ZZ}:
\begin{equation}\label{q5261}
    \dot{q}_i(x)=-\dfrac{1}{h(x)}\varepsilon e^{-\varepsilon p_i(x)}\dfrac{\prod\limits_{j=1}^N\vartheta(q_i(x)-q_j(x-\varepsilon)+\nu\varepsilon)}{\prod\limits_{j:j\neq i}^N\vartheta(q_i(x)-q_j(x))}\,,
\end{equation}
\begin{equation}\label{q5271}
\begin{gathered}
    \varepsilon\dot{p}_i(x)=\sum_{l=1}^N\dot{q}_i(x)E_1(q_i(x)-q_l(x-\varepsilon)+\nu\varepsilon)+\sum_{l=1}^N\dot{q}_{l}(x+\varepsilon)E_1(q_i(x)-q_l(x+\varepsilon)-\nu\varepsilon)-\\
    -\sum_{l:l\neq i}^N(\dot{q_{i}}(x)+\dot{q}_l(x))E_1(q_i(x)-q_l(x))\,.
\end{gathered}
\end{equation}
 It was demonstrated in \cite{ZZ} that these equations of motion can be written in the form of the semi-discrete Zakharov-Shabat equation~\eqref{q19} with the $U$-matrix given by~\eqref{q522} and the $V$-matrix with matrix elements
\begin{equation}\label{q5281}
    \begin{gathered}
        V_{ij}(z,x)=-(1-\delta_{ij})\phi(z,q_i(x)-q_j(x))\dot{q}_j(x)-\delta_{ij}E_1(z)\dot{q}_i(x)+
        \\
        +\delta_{ij}\Bigg(\sum_{m:m\neq i}^N\dot{q}_m(x)E_1(q_i(x)-q_m(x))-
        \sum_{m=1}^N\dot{q}_m(x+\varepsilon)E_1(q_i(x)-q_m(x+\varepsilon)-\nu\varepsilon)\Bigg)\,.
    \end{gathered}
\end{equation}

\subsection{Classical $r$-matrix structure}
The $U$-matrix (\ref{q522}) satisfies the following quadratic r-matrix structure:
\begin{equation}\label{q523}
     \begin{array}{c}
  \displaystyle{
         \left\{U_1(z,x),U_2(w,y)\right\} =
         }
         \\ \ \\
  \displaystyle{
         =
        -\varepsilon\delta(x-y)\Big(U_1(z,x)U_2(w,x){\bf r}_{12}(z-w|x)-
        {\bf r}_{12}(z-w|x-\varepsilon)U_1(z,x)U_2(w,x)+
         }
         \\ \ \\
  \displaystyle{
        +U_1(z,x){\bf s}_{12}^+(z,x)U_2(w,x)-U_2(w,x){\bf s}_{12}^-(w,x)U_1(z,x)\Big)-
         }
         \\ \ \\
  \displaystyle{
        -\varepsilon
        \delta(x-y+\varepsilon)U_1(z,x)U_2(w,y){\bf s}_{12}^-(w,y)+
        \varepsilon\delta(x-y-\varepsilon)U_1(z,x)U_2(w,y){\bf s}_{12}^+(z,x)\,,
        }
  \end{array}
\end{equation}
where ${\bf r}_{12}(z-w|x)$ defined in the same way as in Section~\ref{sec2}:
\begin{equation}\label{q1261}
     \begin{array}{c}
  \displaystyle{
    {\bf r}_{12}(z-w|x) = \sum_{i\neq j}^N\phi(z-w,q_{i}(x)-q_j(x))E_{ij}\otimes E_{ji}+
         }
         \\ \ \\
  \displaystyle{
    +E_1(z-w)\sum_{i=1}^N E_{ii}\otimes E_{ii} - \sum_{i\neq j}^{N} E_1(q_{i}(x)-q_j(x))E_ {ii}\otimes E_{jj}\,,
    }
      \end{array}
\end{equation}
 and
\begin{equation}\label{q524}
    {\bf s}_{12}^{+}(z,x)={\bf s}_{12}(z,x)+{\bf u}_{12}^{+}(x)\,,
\end{equation}
\begin{equation}\label{q525}
    {\bf s}_{12}^{-}(w,x)={\bf s}_{21}(w,x)+{\bf u}_{12}^{-}(x)\,,
\end{equation}
where
\begin{equation}\label{q526}
    {\bf u}_{12}^{+}(x)=\sum_{i,j=1}^NE_1(q_j(x-\varepsilon)-q_i(x)-\nu\varepsilon)E_{ii}\otimes E_{jj}\,,
\end{equation}
\begin{equation}\label{q527}
    {\bf u}_{12}^{-}(x)=\sum_{i,j=1}^NE_1(q_i(x-\varepsilon)-q_j(x)-\nu\varepsilon)E_{ii}\otimes E_{jj}\,,
\end{equation}
\begin{equation}\label{q528}
     \begin{array}{c}
  \displaystyle{
        U_1(z,x){\bf s}_{12}(z,x)=\sum_{i,j=1}^NU_{ij}(z,x)(E_1(z+q_i(x-\varepsilon)-q_j(x)-\nu\varepsilon)-
         }
         \\ \ \\
  \displaystyle{
        -E_1(q_i(x-\varepsilon)-q_j(x)-\nu\varepsilon))E_{ij}\otimes E_{ii}\,,
        }
  \end{array}
\end{equation}
and
\begin{equation}\label{q529}
     \begin{array}{c}
  \displaystyle{
     U_2(w,x){\bf s}_{21}(w,x)=\sum_{i,j=1}^NU_{ij}(w,x)(E_1(w+q_i(x-\varepsilon)-q_j(x)-\nu\varepsilon)-
         }
         \\ \ \\
  \displaystyle{
     -E_1(q_i(x-\varepsilon)-q_j(x)-\nu\varepsilon))E_{ii}\otimes E_{ij}\,.
        }
  \end{array}
\end{equation}
The proof is the same as in the finite-dimensional case.

\section{Non-relativistic limit to Calogero-Moser field theory}\label{sec5}
\setcounter{equation}{0}

\subsection{1+1 Calogero-Moser field theory}
Here we recall main facts on the 1+1 Calogero-Moser field theory \cite{Krich2,LOZ,Krich22}.
Our description coincides with the notations from \cite{Z24}.
\paragraph{Equations of motion.}
The Hamiltonian (\ref{q12})-(\ref{q13}) together with the canonical Poisson brackets (\ref{q11})
provides the following equations of motion:
\begin{equation}
    \dot{q}_i\equiv \left\{H^{\hbox{\tiny{2dCM}}},q_i\right\} = -2(kq_{i,x}+\nu)\left(p_i-\dfrac{1}{N\nu}\sum_{j=1}^Np_j(kq_{j,x}+\nu)\right),
\end{equation}
\begin{equation}
    \begin{gathered}
        \dot{p}_i\equiv \left\{H^{\hbox{\tiny{2dCM}}},p_i\right\}=-k\partial_x\left(p_i^2-2p_i
        \dfrac{1}{N\nu}\sum_{j=1}^Np_j(kq_{j,x}+\nu)+
        \dfrac{1}{2}\dfrac{k^3q_{i,xxx}}{kq_{i,x}+\nu}-\dfrac{1}{4}\dfrac{k^4q^2_{i,xx}}{(kq_{i,x}+\nu)^2}\right)-
        \\
        -2\sum_{j\neq i}^N\left((kq_{j,x}+\nu)^3\wp'(q_{ij})-3k^2(kq_{j,x}+\nu)q_{j,xx}\wp(q_{ij})-k^3q_{j,xxx}\zeta(q_{ij})\right)
    \end{gathered}
\end{equation}
Introduce notations
\begin{equation}\label{q531}
    \alpha_i=(kq_{i,x}+\nu)^{1/2}, \quad i=1,\dots,N
\end{equation}
and
\begin{equation}
    \kappa = -\dfrac{1}{Nc}\sum_{j=1}^Np_{j}(kq_{j,x}+\nu)=-\dfrac{1}{N\nu}\sum_{j=1}^Np_j\alpha_j^2.
\end{equation}
In this notation the Hamiltonian density takes a more compact form:
\begin{equation}
\begin{gathered}
H^{\hbox{\tiny{2dCM}}}(x) = -\sum_{i=1}^Np_i^2\alpha_i^2+N\nu\kappa^2+\sum_{i=1}^Nk^2\alpha_{i,x}^2+
\\
+\dfrac{k}{2}\sum_{i\neq j}^N\left(\alpha_i\alpha_{j,x}-\alpha_j\alpha_{i.x} + \nu(\alpha_{i,x}-\alpha_{j,x})\right)\zeta(q_{ij})+\dfrac{1}{2}\sum_{i\neq j}^N\left(\alpha_i^4\alpha_j^2+\alpha_i^2\alpha_j^4-\nu(\alpha_i^2-\alpha_j^2)^2\right)\wp(q_{ij})
\end{gathered}
\end{equation}
and for the equations of motion we have
\begin{equation}\label{A_EOM_F_CM0}
    \dot{q}_i = -2\alpha_i^2(p_i+\kappa),
\end{equation}
\begin{equation}\label{A_EOM_F_CM}
    \dot{p}_i = -k\partial_x\left(p_i^2+2\kappa p_i+
    k^2\dfrac{\alpha_{i,xx}}{\alpha_i}\right)-2\sum_{j\neq i}^N\left(\alpha_j^6\wp'(q_{ij})-
    6\alpha_j^3\alpha_{j,x}\wp(q_{ij})-k^2\partial_x^2(\alpha_j^2)\zeta(q_{ij})\right)\,.
\end{equation}
%
\paragraph{Zakharov-Shabat equation.}
It was shown in \cite{Krich22} (see also \cite{Z24}, where the same notations as in this paper are used)
that the equations of motion~\eqref{A_EOM_F_CM0}-\eqref{A_EOM_F_CM} are represented in the form of Zakharov-Shabat equations~\eqref{q14}. Introduce the $U-V$ pair:
\begin{equation}\label{F_CM_U}
U_{ij}^{\hbox{\tiny{2dCM}}}(z) = \delta_{ij}\left(p_i+\alpha_i^2E_1(z)-
k\dfrac{\alpha_{i,x}}{\alpha_i}\right)+(1-\delta_{ij})\phi(z,q_{ij})\alpha_j^2,
\end{equation}
\begin{equation}\label{F_CM_V}
\begin{gathered}
    V_{ij}^{\hbox{\tiny{2dCM}}}(z) = \delta_{ij}\left(q_{i,t}E_1(z)-
    N\nu\alpha_i^2\wp(z)-m^0_i-\dfrac{\alpha_{i,t}}{\alpha_i}\right)+\\
    +(1-\delta_{ij})\left(N\nu f(z,q_{ij})-N\nu E_1(z)\phi(z,q_{ij})-m_{ij}\phi(z,q_{ij})\right)\alpha_j^2,
\end{gathered}
\end{equation}
where
\begin{equation}
    m^0_i=p_i^2+2\kappa p_i + k^2\dfrac{\alpha_{i,xx}}{\alpha_i}-\sum_{j\neq i}^N\left((2\alpha_j^4+\alpha_i^2\alpha_j^2)\wp(q_{ij})+4k\alpha_j\alpha_{j,x}\zeta(q_{ij})\right),
\end{equation}
and
\begin{equation}
    m_{ij}=p_i+p_j+2\kappa-k\dfrac{\alpha_{i,x}}{\alpha_i}+k\dfrac{\alpha_j,x}{\alpha_j}+\sum_{l\neq i,j}^N\alpha_l^2\eta(q_i,q_j,q_l), \quad i\neq j,
\end{equation}
with
\begin{equation}
    \eta(z_1,z_2,z_3)=E_1(z_{12})+E_1(z_{23})+E_1(z_{31})\,.
\end{equation}
Then the Zakharov-Shabat equations~\eqref{q14} are equivalent to the equations of motion
identically in $z$.

\paragraph{Classical $r$-matrix.}
It was shown in \cite{Z24} that the $U$-matrix~\eqref{F_CM_U} satisfies the Maillet $r$-matrix structure~\eqref{q16}
known for the finite-dimensional Calogero-Moser model, where positions of particles $q_i$ are replaced with the fields $q_i(x)$:
\beq\label{e69}
  \begin{array}{c}
  \displaystyle{
 {\bf r}^{\hbox{\tiny{2dCM}}}_{12}(z,w|x) =
 (E_1(z-w)+E_1(w))\,\sum\limits_{i=1}^NE_{ii}\otimes E_{ii}+
  }
  \\
   \displaystyle{
+
 \sum\limits^N_{i\neq j}\phi(z-w,q_i(x)-q_j(x))\,E_{ij}\otimes E_{ji}
 -\sum\limits^N_{i\neq j}\phi(-w,q_i(x)-q_j(x))\,E_{ii}\otimes E_{ji}\,.
 }
 \end{array}
 \eq

\subsection{The Maillet $r$-matrix structure from non-relativistic limit}
The aim of this subsection is to
reproduce the Maillet $r$-matrix structure (\ref{q16}) in the continuous non-relativistic limit.
\subsubsection{Limit in the $U$-matrix}
Using the expansion ~\eqref{a06} near $\varepsilon=0$, we have:
\begin{equation}\label{q5211}
\begin{gathered}
    \phi(z,q_i(x-\varepsilon)-q_j(x)-\nu\varepsilon) = \delta_{ij}\left(-\dfrac{1}{\varepsilon\alpha_i^2(x)}+E_1(z)-
    \varepsilon\alpha_i^2(x)\dfrac{E_1^2(z)-\wp(z)}{2}\right)+
    \\+(1-\delta_{ij})\left(\phi(z,q_i(x)-q_j(x))-
    \varepsilon\alpha_i^2(x)f(z,q_{i}(x)-q_{j}(x)\right)+O(\varepsilon^2)
\end{gathered}
\end{equation}
with $\alpha_i^2(x)=\partial_xq_{i}(x)+\nu$ (\ref{q531}). Consider the next factor entering $U$-matrix:
\begin{equation}\label{q5212}
    \dfrac{\prod\limits_{a=1}^N
    \vartheta\left(q_{j}(x)-q_a(x-\varepsilon)+\nu\varepsilon\right)}
    {\vartheta(\nu\varepsilon)\prod\limits_{a\neq j}^N\vartheta(q_j(x)-q_a(x))} = \dfrac{\alpha_j^2(x)}{\nu}
    \left(1+\varepsilon\sum_{a\neq j}^N\alpha_a^2E_1(q_j(x)-q_a(x))\right)+O(\varepsilon^2)\,.
\end{equation}
Finally,
\begin{equation}\label{q52131}
    e^{-\varepsilon p_j(x)} = 1-\varepsilon p_j(x)+O(\varepsilon^2)\,.
\end{equation}
Multiplying all above expressions  we get the following result:
\begin{equation}\label{q5213}
\begin{gathered}
    \nu U^{\varepsilon\to0}_{ij} = -\dfrac{\delta_{ij}}{\varepsilon}+
    \delta_{ij}\left(\tilde{p}_i(x)+E_1(z)\alpha_i^2(x)\right)+
    (1-\delta_{ij})\alpha_j^2(x)\phi(z,q_i(x)-q_j(x))+O(\varepsilon),
\end{gathered}
\end{equation}
where
\begin{equation}\label{q5214}
    \tilde{p}_i(x)=p_i(x)-\sum_{m:m\neq i}^N\alpha_m^2(x)E_1(q_i(x)-q_m(x))
\end{equation}
is a canonical map.
In terms of $U$-matrix for the Calogero-Moser field theory (\ref{F_CM_U}) we have:
\begin{equation}\label{q5215}
    U_{ij}^{\varepsilon\to 0}(z,x)=
    -\dfrac{\delta_{ij}}{\nu\varepsilon}+\dfrac{1}{\nu}U^{2dCM}_{ij}(z,x) +O(\varepsilon)\,.
\end{equation}
\subsubsection{Limit in the $r$-matrix structure}
Now consider the limit in $r$-matrix structure~\eqref{q523}.
The left-hand side takes the form:
\begin{equation}\label{q5221}
    \left\{U_1(z,x),U_2(w,y)\right\}
    \to\dfrac{1}{\nu^2}\left\{U^{\hbox{\tiny{2dCM}}}_1(z,x),U_2^{\hbox{\tiny{2dCM}}}(w,y)\right\}+O(\varepsilon).
\end{equation}
\paragraph{The non-ultralocal terms.}
Consider first the non-ultralocal term in the right-hand side of \eqref{q523}:
\begin{equation}\label{q5222}
    \begin{gathered}
        -\varepsilon
        \delta(x-y+\varepsilon)U_1(z,x)U_2(w,y){\bf s}_{12}^-(w,y)+
        \varepsilon\delta(x-y-\varepsilon)U_1(z,x)U_2(w,y){\bf s}_{12}^+(z,x).
    \end{gathered}
\end{equation}
Taking the limit $\varepsilon \to 0$ we obtain:
\begin{equation}\label{q5223}
    \begin{gathered}
        -\varepsilon\delta(x-y)\left(U_1(z,x)U_2(w,x)s^{-}_{12}(w,x)-U_1(z,x)U_2(w,x)s^+_{12}(z,x)\right)-
        \\-\varepsilon^2\delta'(x-y)\left(U_1(z,x)U_2(w,y){\bf s}_{12}^-(w,y)+U_1(z,x)U_2(w,y)s^+_{12}(z,x)\right).
    \end{gathered}
\end{equation}
Let us focus on the non-ultralocal term with $\delta'(x-y)$. Written in components, it is as follows:
\begin{equation}\label{q5224}
\begin{gathered}
    -\varepsilon^2\delta'(x-y)
    \sum_{i,j,k,l=1}^NU_{ij}(z,x)U_{kl}(w,y)\Big[\delta_{jk}
    E_1(w+q_k(y-\varepsilon)-q_l(y)-\nu\varepsilon)+
    \\
    +\delta_{il}E_1(z+q_i(x-\varepsilon)-q_j(x)-\nu\varepsilon)+
    (1-\delta_{jk})E_1(q_j(y-\varepsilon)-q_l(y)-\nu\varepsilon)+
    \\
    +(1-\delta_{il})E_1(q_l(x-\varepsilon)-q_j(x)-\nu\varepsilon)\Big]E_{ij}\otimes E_{kl}.
\end{gathered}
\end{equation}
In what follows we use the following simple expansions:
\begin{equation}\label{q5225}
    \begin{gathered}
        E_1(w+q_k(x-\varepsilon)-q_l(x)-\nu\varepsilon)\to E_1(w+q_k(x)-q_l(x))+O(\varepsilon),
        \\
        E_1(q_k(x-\varepsilon)-q_l(x)-\nu\varepsilon)\to -\dfrac{\delta_{kl}}{\varepsilon\alpha_{k}^2(x)}+(1-\delta_{kl})E_1(q_k(x)-q_l(x))+O(\varepsilon).
    \end{gathered}
\end{equation}
By taking the limit in (\ref{q5224}) we get:
\begin{equation}\label{q5226}
\begin{gathered}
    -\delta'(x-y)\dfrac{1}{\nu^2}\big((E_1(z)+E_1(w))\sum_{i=1}^NE_{ii}\otimes E_{ii}+\sum_{i\neq j=1}^N\phi(w,q_{ji}(y))E_{ii}\otimes E_{ji}+
    \\
    + \sum_{i\neq j=1}^N\phi(z,q_{ij}(x))E_{ij}\otimes E_{jj}\big).
\end{gathered}
\end{equation}
Next, we use the relation
\begin{equation}\label{q5227}
    \delta'(x-y)f(y)=\delta'(x-y)f(x)+\delta(x-y)f'(x)\,.
\end{equation}
Ignoring local terms for now, and summing up all
the terms with $\delta'(x-y)$, we obtain the non-ultralocal part in the form:
\begin{equation}\label{q5228}
\begin{gathered}
    -\dfrac{1}{\nu^2}\delta'(x-y)\big((E_1(z)+E_1(w))\sum_{i=1}^NE_{ii}\otimes E_{ii}+\sum_{i\neq j=1}^N\phi(w,q_{ji}(x))E_{ii}\otimes E_{ji}+
    \\
    + \sum_{i\neq j=1}^N\phi(z,q_{ij}(x))E_{ij}\otimes E_{jj}\big)=-\dfrac{1}{\nu^2}\delta'(x-y)
    \left(\mathbf{r}_{12}^{\hbox{\tiny{2dCM}}}(z,w|x)+\mathbf{r}_{21}^{\hbox{\tiny{2dCM}}}(w,z|x)\right).
\end{gathered}
\end{equation}
This part completely coincides with the
non-ultralocal part of the Maillet $r$-matrix structure (\ref{q16}).
\paragraph{The ultralocal terms.}
We are now left with the ultralocal terms.
Aside from the local part of~\eqref{q523},
we also got extra terms from the non-ultralocal part.
After summing up all these terms together we get the local term in the following form:
\begin{equation}\label{q5229}
    \begin{gathered}
        -\delta(x-y)\Bigg[\dfrac{1}{\nu^2}\sum_{i\neq j=1}^N(\alpha_j^2(x)-
        \alpha_i^2(x))f(w,q_{ji}(x))E_{ii}\otimes E_{ji}\Bigg]-\\-\delta(x-y)
        \lim_{\varepsilon\to 0}\varepsilon\Big[U_1(z,x)U_2(w,x){\bf r}_{12}(z-w|x)-
        {\bf r}_{12}(z-w|x-\varepsilon)U_1(z,x)U_2(w,x)+\\+\left[U_1(z,x){\bf s}_{12}^+(z,x),U_2(w,x)\right]-
        \left[U_2(w,x){\bf s}_{12}^-(w,x),U_1(z,x)\right]\Big].
    \end{gathered}
\end{equation}
In order to prove that we get the Maillet $r$-matrix structure in the limit, one should prove the relation
\begin{equation}\label{q52210}
    \begin{gathered}
        0=-\partial_x\mathbf{r}^{\hbox{\tiny{2dCM}}}_{12}(z,w|x)+
        \left[U_1^{\hbox{\tiny{2dCM}}}(z,x),\mathbf{r}^{\hbox{\tiny{2dCM}}}_{12}(z,w|x)\right]-\left[U_2^{\hbox{\tiny{2dCM}}}(w,x), \mathbf{r}^{\hbox{\tiny{2dCM}}}_{21}(w,z|x)\right]+
        \\
        +\Bigg[\sum_{i\neq j=1}^N(\alpha_j^2(x)-\alpha_i^2(x))f(w,q_{ji}(x))E_{ii}\otimes E_{ji}\Bigg]+
        \\
        +\lim_{\varepsilon\to 0}\varepsilon\Big[U_1(z,x)U_2(w,x){\bf r}_{12}(z-w|x)-
        {\bf r}_{12}(z-w|x-\varepsilon)U_1(z,x)U_2(w,x)+\\+\left[U_1(z,x){\bf s}_{12}^+(z,x),U_2(w,x)\right]-
        \left[U_2(w,x){\bf s}_{12}^-(w,x),U_1(z,x)\right]\Big]\,.
    \end{gathered}
\end{equation}
We prove this relation in the Appendix C.

%
%
%
%
%

\section{Appendix A: elliptic functions}
\def\theequation{A.\arabic{equation}}
\setcounter{equation}{0}

We mainly deal with the elliptic Kronecker function
 \beq\label{a01}
  \begin{array}{l}
  \displaystyle{
 \phi(z,u)=\frac{\vth'(0)\vth(z+u)}{\vth(z)\vth(u)}\,,
 }
 \end{array}
 \eq
 where $\vth(z)$ is the first Jacobi theta-function. In Riemann's notation
 it is as follows.
 Define the theta-functions with characteristics $a,b$:
\beq\label{a02}
 \begin{array}{c}
  \displaystyle{
\theta{\left[\begin{array}{c}
a\\
b
\end{array}
\right]}(z|\, \tau ) =\sum_{j\in \z}
\exp\left(2\pi\imath(j+a)^2\frac\tau2+2\pi\imath(j+a)(z+b)\right)\,,\quad {\rm Im}(\tau)>0\,,
}
 \end{array}
 \eq
where $a\,,b\in\frac{1}{N}\,\mZ$.
 In particular, the odd theta function $\vth(z)$ ($\theta_1(z)$ in the Jacobi notation) is
 \beq\label{a03}
 \begin{array}{c}
  \displaystyle{
\vth(z)=\vth(z,\tau)\equiv-\theta{\left[\begin{array}{c}
1/2\\
1/2
\end{array}
\right]}(z|\, \tau )\,.
 }
 \end{array}
 \eq
 The Kronecker function has a single simple pole in variable $z$ at $z=0$:
 \beq\label{a04}
  \begin{array}{l}
  \displaystyle{
\res\limits_{z=0}\phi(z,u)=1
 }
 \end{array}
 \eq
 The following quasi-periodicity properties hold:
 \beq\label{a05}
  \begin{array}{l}
  \displaystyle{
 \phi(z+1,u)= \phi(z,u)\,,\qquad  \phi(z+\tau,u)= \exp (-2\pi\imath u)\phi(z,u)\,.
 }
 \end{array}
 \eq
 The expansion near $z=0$ has the form
 \beq\label{a06}
  \begin{array}{l}
  \displaystyle{
 \phi(z,u)=\frac{1}{z}+E_1(u)+\frac{E_1^2(u)-\wp(u)}{2}+O(z^2),
 }
 \end{array}
 \eq
 where
 \beq\label{a07}
  \begin{array}{l}
  \displaystyle{
 E_1(u)=\frac{\vth'(u)}{\vth(u)}
 }
 \end{array}
 \eq
is the first Eisenstein function.
The relation to the Weierstrass functions is as follows:
\beq\label{a071}
\displaystyle{
\zeta(z)=E_1(z)+2\eta_0 z\,,\quad \eta_0=-\frac{1}{6}\frac{\vth'''(0)}{\vth'(0)}\,,\quad
\sigma(z)=\frac{\vth(z)}{\vth'(0)}\,e^{\eta_0 z^2}
}
\eq
and
 \beq\label{a072}
  \begin{array}{l}
  \displaystyle{
 \wp(u)=E_2(u)+\frac{1}{3}\frac{\vth'''(0)}{\vth'(0)}\,,
 }
 \end{array}
 \eq
 where $E_2(u)$ is the second Eisenstein function
 \beq\label{a073}
  \begin{array}{l}
  \displaystyle{
 E_2(u)=-E_1'(u)=-\p^2_u\log\vth(u)\,.
 }
 \end{array}
 \eq
 The parities of the functions are
 \beq\label{a074}
  \begin{array}{l}
  \displaystyle{
 \phi(z,u)=-\phi(-z,-u)\,,\quad E_1(u)=-E_1(-u)\,,\quad E_2(u)=E_2(-u)\,.
 }
 \end{array}
 \eq

It follows from the definition (\ref{a09}) that
 \beq\label{a08}
  \begin{array}{l}
  \displaystyle{
 \p_z\phi(z,u)=(E_1(z+u)-E_1(z))\phi(z,u)\,,
 }
 \\ \ \\
   \displaystyle{
  \p_u\phi(z,u)=(E_1(z+u)-E_1(u))\phi(z,u)\,.
 }
 \end{array}
 \eq
A set of the widely known addition formulae (the genus one Fay identity and its degenerations) is used in this paper:
\beq\label{a10}
  \begin{array}{c}
  \displaystyle{
  \phi(z_1, u_1) \phi(z_2, u_2) = \phi(z_1, u_1 + u_2) \phi(z_2 - z_1, u_2) + \phi(z_2, u_1 + u_2) \phi(z_1 - z_2, u_1)
 }
 \end{array}
 \eq
\beq\label{a09}
  \begin{array}{c}
  \displaystyle{
 \phi(z,u_1)\phi(z,u_2)=\phi(z,u_1+u_2)\Big(E_1(z)+E_1(u_1)+E_1(u_2)-E_1(z+u_1+u_2)\Big)\,,
 }
 \end{array}
 \eq
and
\beq\label{a11}
  \begin{array}{c}
  \displaystyle{
  \phi(z,u_1-v)\phi(w,u_2+v)\phi(z-w,v)-\phi(z,u_2+v)\phi(w,u_1-v)\phi(z-w,u_1-u_2-v)=
 }
 \\ \ \\
   \displaystyle{
  =\phi(z,u_1)\phi(w,u_2)\Big( E_1(v)-E_1(u_1-u_2-v)+E_1(u_1-v)-E_1(u_2+v) \Big)\,.
 }
 \end{array}
 \eq
Also,
\beq\label{a12}
  \begin{array}{c}
  \displaystyle{
 \phi(z,u)\phi(z,-u)=\wp(z)-\wp(u)=E_2(z)-E_2(u)\,.
 }
 \end{array}
 \eq

\section{Appendix B: classical $r$-matrix for Ruijsenaars-Schneider model}
\def\theequation{B.\arabic{equation}}
\setcounter{equation}{0}
\paragraph{Notations.} The Ruijsenaars-Schneider model is described by the quadratic $r$-matrix structure of the following
type \cite{FM,STS,Suris}:
\beq\label{b01}
  \begin{array}{c}
  \displaystyle{
\left\{L_1(z),L_2(w)\right\} =
 }
 \\ \ \\
   \displaystyle{
a_{12}(z,w)L_1(z)L_2(w)+L_1(z)b_{12}(z,w)L_2(w)
-L_2(z)c_{12}(z,w)L_1(w)-L_1(z)L_2(w)d_{12}(z,w)\,,
 }
 \end{array}
 \eq
where
 $a_{12}(z,w),b_{12}(z,w),c_{12}(z,w),d_{12}(z,w)$ are $\Mat^{\otimes 2}$-valued functions on
the phase space depending also on the spectral parameters $z,w\in\mC$.
Here we assume the standard notations $L_1(z)=L(z)\otimes 1_N$, $L_2(w)=1_N\otimes L(w)$, $1_N$ -- $N\times N$ identity matrix.
The l.h.s. of (\ref{b01}) is by definition
\beq\label{b02}
  \begin{array}{c}
  \displaystyle{
\left\{L_1(z),L_2(w)\right\} =\sum\limits_{i,j,k,l=1}^NE_{ij}\otimes E_{kl}\{L_{ij}(z),L_{kl}(w)\}\,,
 }
 \end{array}
 \eq
where $E_{ij}$ are matrix units with elements $(E_{ij})_{ab}=\delta_{ia}\delta_{jb}$.
It is assumed that (\ref{b01}) holds true identically in variables  $z$ and $w$.
Changing the indices 1,2 in any expression means interchanging of tensor components
\beq\label{b03}
  \begin{array}{c}
  \displaystyle{
\left\{L_2(w),L_1(z)\right\} =\sum\limits_{i,j,k,l=1}^NE_{kl}\otimes E_{ij}\{L_{ij}(z),L_{kl}(w)\}=
P_{12}\left\{L_1(z),L_2(w)\right\}P_{12}\,,
 }
 \end{array}
 \eq
where $P_{12}$ is the matrix permutation operator
\beq\label{b04}
  \begin{array}{c}
  \displaystyle{
P_{12}=\sum\limits_{i,j=1}^NE_{ij}\otimes E_{ji}\in\Mat^{\otimes 2}\,.
 }
 \end{array}
 \eq
It has the property $P_{12}^2=1_{N}\otimes 1_N=1_{N^2}$ and $P_{12}(A\otimes B)=(B\otimes A)P_{12}$
for any $A,B\in\Mat$.

The skew-symmetry of the Poisson bracket $\left\{L_1(z),L_2(w)\right\}=-\left\{L_2(w),L_1(z)\right\}$
provides conditions
\beq\label{b05}
  \begin{array}{c}
  \displaystyle{
a_{12}(z,w)=-a_{21}(w,z)\,,
\quad
c_{12}(z,w)=b_{21}(w,z)\,,\quad
d_{12}(z,w)=-d_{21}(w,z)\,,
 }
 \end{array}
 \eq
A sufficient condition for the Poisson commutativity
\beq\label{b06}
  \begin{array}{c}
  \displaystyle{
\left\{\tr L^k(z),\tr L^m(w)\right\} =0
}
 \end{array}
 \eq
is
\beq\label{b07}
  \begin{array}{c}
  \displaystyle{
a_{12}(z,w)+b_{12}(z,w)=c_{12}(z,w)+d_{12}(z,w)\,.
}
 \end{array}
 \eq

\paragraph{$r$-matrix from \cite{NKSR}.}
Introduce the Lax matrix
  \beq\label{b08}
  \begin{array}{c}
  \displaystyle{
 \tiL^{\hbox{\tiny{RS}}}_{ij}(z)=b_i\phi(z,q_{i}-q_j+\eta)\,,\ i,j=1,\ldots ,N\,,
 }
 \end{array}
 \eq
where $b_i$ is given by (\ref{q102}). It differs from the Lax matrix used in  \cite{NKSR} by only
common factor $\sigma'(0)$ and usage of the Weierstrass sigma function (\ref{a071}) instead of
theta function. All required identities (\ref{a08})-(\ref{a11}) hold true for the definition with
sigma function as well, but the Eisenstein function $E_1(z)$ should be replaced with the
Weierstrass zeta function $\zeta(z)$. These two descriptions are related by a simple gauge transformation.

The Lax matrix (\ref{b08}) satisfies (\ref{b01}) with
  \beq\label{b09}
  \begin{array}{c}
  \displaystyle{
 a_{12}(z,w)=-{\ti r}^+_{12}(z,w)\,,\quad d_{12}(z,w)=-{\ti r}^-_{12}(z,w)\,,
 }
 \end{array}
 \eq
  \beq\label{b10}
  \begin{array}{c}
  \displaystyle{
 b_{12}(z,w)={\ti s}^+_{12}(z,w)\,,\quad c_{12}(z,w)={\ti s}^-_{12}(z,w)\,,
 }
 \end{array}
 \eq
where
\begin{equation}\label{b11}
   \displaystyle{
    {\ti r}^-_{12}(z,w) = {\ti r}_{12}(z,w)-{\ti s}_{12}(z)+{\ti s}_{21}(w)\,,
    }
\end{equation}
\begin{equation}\label{b12}
    {\ti r}_{12}^+(z,w) = r_{12}(z,w)+u^+_{12}+u^-_{12}\,,
\end{equation}
\begin{equation}\label{b13}
    {\ti s}_{12}^+ (z,w) = {\ti s}_{12}(z)+u_{12}^+\,,
\end{equation}
\begin{equation}\label{b14}
    {\ti s}_{12}^- (z,w) = {\ti s}_{21}(w)-u_{12}^-
\end{equation}
with
\begin{equation}\label{b15}
    {\ti r}_{12}(z,w) = \sum_{i\neq j}^N\phi(z-w,q_{ij})E_{ij}\otimes E_{ji}+E_1(z-w)\sum_{i=1}^N E_{ii}\otimes E_{ii} + \sum_{i\neq j}^{N} E_1(q_{ij})E_ {ii}\otimes E_{jj}\,,
\end{equation}
\begin{equation}\label{b16}
    {\ti s}_{12}(z) = \sum_{i,j=1}^N \Big(\tiL^{\hbox{\tiny{RS}}}(z)^{-1}\,''\partial_{\eta}\tiL^{\hbox{\tiny{RS}}}(z)''\Big)_{ij}\, E_{ij}\otimes E_{jj}
\end{equation}
and $u_{12}^\pm$ from (\ref{q127}). The
notation $''\partial_{\eta}\tiL^{\hbox{\tiny{RS}}}(z)''$ is as in (\ref{q129}):
\begin{equation}\label{b17}
''\partial_{\eta}\tiL^{\hbox{\tiny{RS}}}(z)''_{ij}=\tiL^{\hbox{\tiny{RS}}}_{ij}(z)\Big(E_1(z+q_{ij}+\eta)-E_1(q_{ij}+\eta)\Big)\,.
\end{equation}
Then
\begin{equation}\label{b18}
    \tiL^{\hbox{\tiny{RS}}}_1(z){\ti s}_{12}(z) = \sum_{i,j=1}^N \tiL^{\hbox{\tiny{RS}}}_{ij}(z)
    \Big(E_1(z+q_{ij}+\eta)-E_1(q_{ij}+\eta)\Big)E_{ij}\otimes E_{ii}\,,
\end{equation}
and in the same way
\begin{equation}\label{b19}
    \tiL^{\hbox{\tiny{RS}}}_2(w){\ti s}_{21}(w) = \sum_{i,j=1}^N
    \tiL^{\hbox{\tiny{RS}}}_{ij}(w)\Big(E_1(w+q_{ij}+\eta)-E_1(q_{ij}+\eta)\Big)E_{jj}\otimes E_{ij}\,.
\end{equation}

\section{Appendix C: limit of the field $r$-matrix structure}
\def\theequation{C.\arabic{equation}}
\setcounter{equation}{0}
Here we prove the relation (\ref{q52210}).

The expression ${\bf r}_{12}(z-w|x)$ is independent of $\varepsilon$
 and ${\bf r}_{12}(z-w|x-\varepsilon)={\bf r}_{12}(z-w|x)-\varepsilon \partial_x{\bf r}_{12}(z-w|x)+O(\varepsilon^2)$,
 so that the terms in the above expression have the form
\begin{equation}\label{q52211}
\begin{gathered}
     -\varepsilon\left(U_1(z,x)U_2(w,x){\bf r}_{12}(z-w|x)-{\bf r}_{12}(z-w|x-\varepsilon)U_1(z,x)U_2(w,x)\right)\to\\ \dfrac{1}{\nu^2}\left(-\partial_x{\bf r}_{12}(z-w|x) +\left[U_1^{\hbox{\tiny{2dCM}}}(z,x),{\bf r}_{12}(z-w|x)\right]+
     \left[U_2^{\hbox{\tiny{2dCM}}}(w,x),{\bf r}_{12}(z-w|x)\right]\right)+O(\varepsilon)\,.
\end{gathered}
\end{equation}
The expression~\eqref{q52210} transforms into
\begin{equation}\label{q52212}
    \begin{gathered}
        0=-\partial_x\left(\mathbf{r}^{\hbox{\tiny{2dCM}}}_{12}(z,w|x)-{\bf r}_{12}(z-w|x)\right)+
        \left[U_1^{\hbox{\tiny{2dCM}}}(z,x),\left(\mathbf{r}^{\hbox{\tiny{2dCM}}}_{12}(z,w|x)-{\bf r}_{12}(z-w|x)\right)\right]-
        \\
        -\left[U_2^{\hbox{\tiny{2dCM}}}(w,x), \left(\mathbf{r}^{\hbox{\tiny{2dCM}}}_{21}(w,z|x) +{\bf r}_{12}(z-w|x)\right)\right]+\\+
        \Bigg[\sum_{i\neq j=1}^N(\alpha_j^2(x)-\alpha_i^2(x))f(w,q_{ji}(x))E_{ii}\otimes E_{ji}\Bigg]+
        \\
        +\nu^2\lim_{\varepsilon\to 0}\varepsilon\Big(\left[U_1(z,x){\bf s}_{12}^+(z,x),U_2(w,x)\right]-
        \left[U_2(w,x){\bf s}_{12}^-(w,x),U_1(z,x)\right]\Big).
    \end{gathered}
\end{equation}
Let us work with different terms independently.
\begin{equation}\label{q52213}
    \begin{gathered}
        -\partial_x\left(\mathbf{r}^{\hbox{\tiny{2dCM}}}_{12}(z,w|x)-{\bf r}_{12}(z-w|x)\right)=-\sum_{i\neq j=1}^N(\alpha_j^2(x)-\alpha_i^2(x))f(w,q_{ji}(x))E_{ii}\otimes E_{ji}-\\-\sum_{i\neq j=1}^N(\alpha_{j}^2(x)-\alpha_i^2(x))E_2(q_{ji}(x))E_{ii}\otimes E_{jj}.
    \end{gathered}
\end{equation}
The first term is cancelled out with the same one in~\eqref{q52212}. The next term gives:
\begin{equation}\label{q52214}
\begin{gathered}
    \left[U_1^{\hbox{\tiny{2dCM}}}(z,x),\left(\mathbf{r}^{\hbox{\tiny{2dCM}}}_{12}(z,w|x)-{\bf r}_{12}(z-w|x)\right)\right]=
    \\
    =E_1(w)\sum_{i,j=1}^NU_{ij}^{\hbox{\tiny{2dCM}}}(z,x)
    E_{ij}\otimes E_{jj}-E_1(w)\sum_{i,j=1}^NU_{ij}^{\hbox{\tiny{2dCM}}}(z,x)E_{ij}\otimes E_{ii}+
    \\
    +\sum_{i,j\neq k=1}^NU_{ij}^{\hbox{\tiny{2dCM}}}(z,x)\phi(w,q_{kj}(x))E_{ij}\otimes E_{kj}-\sum_{j,i\neq k=1}^NU_{ij}^{\hbox{\tiny{2dCM}}}(z,x)\phi(w,q_{ki}(x))E_{ij}\otimes E_{ki}+
    \\
    -\sum_{i,j\neq k=1}^NU_{ij}^{\hbox{\tiny{2dCM}}}(z,x)E_1(q_{kj}(x))E_{ij}\otimes E_{kk}+\sum_{j,i\neq k=1}^NU_{ij}^{\hbox{\tiny{2dCM}}}(z,x)E_1(q_{ki}(x))E_{ij}\otimes E_{kk}
\end{gathered}
\end{equation}
and
\begin{equation}\label{q_52215}
    \begin{gathered}
        -\left[U_2^{\hbox{\tiny{2dCM}}}(w,x), \left(\mathbf{r}^{\hbox{\tiny{2dCM}}}_{21}(w,z|x+{\bf r}_{12}(z-w|x)\right)\right]=
        \\
        =E_1(z)\sum_{i,j=1}^NU_{ij}^{\hbox{\tiny{2dCM}}}(w,x)
        E_{ii}\otimes E_{ij}-E_1(z)\sum_{i,j=1}^NU_{ij}^{\hbox{\tiny{2dCM}}}(w,x)E_{jj}\otimes E_{ij}+
        \\
        +\sum_{j,i\neq k=1}^NU_{ij}^{\hbox{\tiny{2dCM}}}(w,x)\phi(z,q_{ki})E_{ki}\otimes E_{ij}-\sum_{i,j\neq k=1}^NU_{ij}^{\hbox{\tiny{2dCM}}}(w,x)\phi(z,q_{kj})E_{kj}\otimes E_{ij}+
        \\
        +\sum_{j,i\neq k=1}^NU_{ij}^{\hbox{\tiny{2dCM}}}(w,x)E_1(q_{ik})E_{kk}\otimes E_{ij}-\sum_{i,j\neq k=1}^NU_{ij}^{\hbox{\tiny{2dCM}}}(w,x)E_1(q_{jk})E_{kk}\otimes E_{ij}\,.
    \end{gathered}
\end{equation}
Now we are left with only the last part of ~\eqref{q52212}:
\begin{equation}\label{q52216}
         \nu^2\varepsilon\left[U_1(z,x){\bf s}_{12}^+(z,x),U_2(w,x)\right] -\nu^2\varepsilon\left[U_2(w,x){\bf s}_{12}^-(w,x),U_1(z,x)\right].
\end{equation}
In components, one has
\begin{equation}\label{q52217}
    \begin{gathered}
       \varepsilon\nu^2 \sum_{i,j,k,l=1}^N
       U_{ij}(z,x)U_{kl}(w,x)E_{ij}\otimes E_{kl}\times
       \\
       \times\Big[\delta_{ik}
       \big(E_1(z+q_i(x-\varepsilon)-q_j(x)-\nu\varepsilon)-
       E_1(w+q_i(x-\varepsilon)-q_l(x)-\nu\varepsilon)\big)+
       \\
       +(1-\delta_{ik})\big(E_1(q_k(x-\varepsilon)-q_j(x)-\nu\varepsilon)-
       E_1(q_i(x-\varepsilon)-q_l(x)-\nu\varepsilon)\big)-
       \\
       -
        \delta_{il}E_1(z+q_i(x-\varepsilon)-q_j(x)-\nu\varepsilon)-
        (1-\delta_{il})E_1(q_l(x-\varepsilon)-q_j(x)-\nu\varepsilon)+
        \\
        +\delta_{jk}E_1(w+q_k(x-\varepsilon)-q_l(x)-\nu\varepsilon)+
        (1-\delta_{jk})E_1(q_j(x-\varepsilon)-q_l(x)-\nu\varepsilon)
        \Big].
    \end{gathered}
\end{equation}
The limit of this expression cancels all the terms which are left in~\eqref{q52212}
after substituting everything in components.

Let us calculate the limit of every component of~\eqref{q52217}.
 It was previously proved that every singular in $\varepsilon$
 term is cancelled out. Therefore, we are looking for only those terms which are independent of $\varepsilon$.
 Let us write this as "$\approx$".
The first term gives
\begin{equation}\label{q52218}
\begin{gathered}
    \varepsilon\nu^2\delta_{ik}U_{ij}(z,x)U_{il}(w,x)
    \Big(E_1(z+q_i(x-\varepsilon)-q_j(x)-\nu\varepsilon)-
    \\-E_1(w+q_i(x-\varepsilon)-q_l(x)-\nu\varepsilon)\Big)E_{ij}\otimes E_{il}\approx
    \\
    -\alpha_i^2(E_2(w)-E_2(z))E_{ii}\otimes E_{ii}-
    \\-U_{ij}^{\hbox{\tiny{2dCM}}}(w,x)(E_1(z)-E_1(w+q_{ij}))E_{ii}\otimes E_{ij}
    -\\-U_{ij}^{\hbox{\tiny{2dCM}}}(z,x)(E_1(z+q_{ij})-E_1(w))E_{ij}\otimes E_{ii}+O(\varepsilon).
\end{gathered}
\end{equation}
The next one yields
\begin{equation}\label{q52219}
\begin{gathered}
    \varepsilon\nu^2(1-\delta_{ik})U_{ij}(z,x)U_{kl}(w,x)
    \Big(E_1(q_k(x-\varepsilon)-q_j(x)-\nu\varepsilon)-
    \\E_1(q_i(x-\varepsilon)-q_l(x)-\nu\varepsilon)\Big)E_{ij}\otimes E_{kl}\approx\\
    -(1-\delta_{ki})U^{\hbox{\tiny{2dCM}}}_{ij}(w,x)\phi(z,q_{ki})E_{ki}\otimes E_{ij}+\\+(1-\delta_{ki})U_{ij}^{\hbox{\tiny{2dCM}}}(z,x)\phi(w,q_{ki})E_{ij}\otimes E_{ki}-\\
    -(1-\delta_{ik})(1-\delta_{jk})U_{ij}^{\hbox{\tiny{2dCM}}}(z,x)E_1(q_{kj})E_{ij}\otimes E_{kk}+\\+(1-\delta_{ki})(1-\delta_{kj})U_{ij}^{\hbox{\tiny{2dCM}}}(w,x)E_1(q_{kj})E_{kk}\otimes E_{ij}-\\
    -(1-\delta_{ki})U_{ij}^{\hbox{\tiny{2dCM}}}(w,x)E_1(q_{ik})E_{kk}\otimes E_{ij}+\\+(1-\delta_{ki})U_{ij}^{\hbox{\tiny{2dCM}}}(z,x)E_1(q_{ik})E_{ij}\otimes E_{kk}-\\
    -(1-\delta_{ij})E_2(q_{ji})(\alpha_i^2-\alpha_j^2)E_{ii}\otimes E_{jj} + O(\varepsilon).
\end{gathered}
\end{equation}
In the same way
\begin{equation}\label{q52220}
    \begin{gathered}
        -\varepsilon\nu^2\delta_{il}
         U_{ij}(z,x)U_{ki}(w,x)E_1(z+q_i(x)-q_j(x+\varepsilon)-\nu\varepsilon)E_{ij}\otimes E_{ki}\approx \\
        -\alpha_i^2E_2(z)E_{ii}\otimes E_{ii}+E_1(z)U^{\hbox{\tiny{2dCM}}}_{ij}(w,x)E_{jj}\otimes E_{ij}+E_1(z+q_{ij})U_{ij}^{\hbox{\tiny{2dCM}}}(z,x)E_{ij}\otimes E_{ii}\,,
    \end{gathered}
\end{equation}
\begin{equation}\label{q52221}
    \begin{gathered}
        -\varepsilon\nu^2(1-\delta_{in})
        U_{ij}(z,x)U_{mn}(w,x)E_1(q_n(x)-q_j(x+\varepsilon)-\nu\varepsilon)E_{ij}\otimes E_{mn}\approx\\
        (1-\delta_{kj})\phi(z,q_{kj})U_{ij}^{\hbox{\tiny{2dCM}}}(w,x)E_{kj}\otimes E_{ij}+(1-\delta_{ik})(1-\delta_{jk})U_{ij}^{\hbox{\tiny{2dCM}}}(z,x)E_1(q_{kj})E_{ij}\otimes E_{kk}+\\+
        (1-\delta_{kj})U_{ij}^{\hbox{\tiny{2dCM}}}(w,x)
        E_1(q_{jk})E_{kk}\otimes E_{ij}-(1-\delta_{ij})\alpha_i^2E_2(q_{ji})E_{ii}\otimes E_{jj}\,,
    \end{gathered}
\end{equation}
\begin{equation}\label{q52222}
    \begin{gathered}
        \varepsilon\nu^2 \delta_{jk} U_{ij}(z,x)U_{kl}(w,x)
        E_1(w+q_k(x-\varepsilon)-q_l(x)-\nu\varepsilon)E_{ij}\otimes E_{kl}\approx\\
        E_2(w)\alpha_i^2E_{ii}\otimes E_{ii}-U_{ij}^{\hbox{\tiny{2dCM}}}(z,x)E_1(w)E_{ij}\otimes E_{jj}-U_{ij}^{\hbox{\tiny{2dCM}}}(w,x)E_1(w+q_{ij})E_{ii}\otimes E_{ij}
    \end{gathered}
\end{equation}
and, finally,
\begin{equation}\label{q52223}
    \begin{gathered}
        \varepsilon\nu^2 (1-\delta_{jk})U_{ij}(z,x)U_{kl}(w,x)
        E_1(q_j(x-\varepsilon)-q_l(x)-\nu\varepsilon)E_{ij}\otimes E_{kl}\approx\\
        -(1-\delta_{kj})U_{ij}^{\hbox{\tiny{2dCM}}}(z,x)\phi(w,q_{kj})E_{ij}\otimes E_{kj}+(1-\delta_{ik})\alpha_i^2E_2(q_{ik})E_{ii}\otimes E_{kk}-\\
        -(1-\delta_{ki})(1-\delta_{kj})U_{ij}^{\hbox{\tiny{2dCM}}}(w,x)E_1(q_{kj})E_{kk}\otimes E_{ij}+(1-\delta_{kj})U_{ij}^{\hbox{\tiny{2dCM}}}(z,x)E_1(q_{kj})E_{ij}\otimes E_{kk}\,.
    \end{gathered}
\end{equation}
Summing up all the terms we conclude that the non-relativistic limit
of the classical $r$-matrix structure of the Ruijsenaars-Schneider field theory~\eqref{q523},
indeed transforms into the Maillet $r$-matrix structure for 1+1 Calogero-Moser theory.
%
%


\paragraph{Acknowledgments.}
We are grateful to A. Zabrodin for useful discussions.
The work of A. Zotov was performed at the Steklov International Mathematical Center and supported by the Ministry of Science and Higher Education of the Russian Federation (agreement no. 075-15-2025-303).





\begin{small}

\end{small}

\end{document}